\documentclass[%
 twocolumn,
 reprint,
 superscriptaddress,
 nofootinbib,
 amsmath,
 aps, pra,
]{revtex4-2}

\usepackage{color}
\usepackage{graphicx}
\usepackage{hyperref}
\usepackage{comment}
           
\usepackage{siunitx}
\renewcommand{\Re}{\mathop{\mathrm{Re}}\nolimits}           
   
\DeclareMathOperator{\Tr}{\text{Tr}}

\begin{document}

\title{Theory of quasiparticle-induced errors in driven-dissipative Schr\"odinger cat qubits}

\author{Kirill S. Dubovitskii}
\affiliation{Univ.~Grenoble Alpes, CNRS, LPMMC, 38000 Grenoble, France}

\author{Denis M. Basko}
\affiliation{Univ.~Grenoble Alpes, CNRS, LPMMC, 38000 Grenoble, France}

\author{Julia S. Meyer}
\affiliation{Univ.~Grenoble Alpes, CEA, Grenoble INP, IRIG, Pheliqs, 38000 Grenoble, France}

\author{Manuel Houzet}
\affiliation{Univ.~Grenoble Alpes, CEA, Grenoble INP, IRIG, Pheliqs, 38000 Grenoble, France}

\begin{abstract}
Understanding the mechanisms of qubit decoherence is a crucial prerequisite for improving the qubit performance. In this paper we discuss the effects of residual Bogolyubov quasiparticles in Schrödinger cat qubits, either of the dissipative or Kerr type. The major difference from previous studies of quasiparticles in superconducting qubits is that the Schrödinger cat qubits are operated under non-equilibrium conditions. 
Indeed, an external microwave drive is needed to stabilize ``cat states'', which are superpositions of coherent degenerate eigenstates of an effective stationary Lindbladian in the rotating frame. 
We present a microscopic derivation of the master equation for cat qubits and express the effect of the quasiparticles as dissipators acting on the density matrix of the cat qubit. This enables us to determine the conditions under which the quasiparticles give a substantial contribution to the qubit errors. 
\end{abstract}

\maketitle

\section{Introduction}

Superconducting circuits represent one of the most promising physical platforms for realizing qubits, the elementary building blocks of quantum computers~\cite{Arute2019}. Operational superconducting qubits include 
transmon \cite{Koch2007}, fluxonium \cite{Manucharyan2009}, and many others. All qubits are subject to errors due to their environment, and quantum error correction imposes a huge overhead cost in any quantum computer architecture, since a single logical qubit must be represented by many physical qubits~\cite{NielsenChuang}. Then, qubits with intrinsic protection against some errors may reduce this cost and offer a technological advantage. One way to implement such intrinsic protection is to encode the qubit states in a bosonic degree of freedom, well separating the two states in the phase space, thus reducing their sensitivity to local noise~\cite{Leghtas2013,Mirrahimi2014}.  This separation can be achieved via an interplay between a microwave drive and non-linear couplings; such Schr\"odinger cat qubits have been successfully fabricated in recent years~\cite{Leghtas2015, Touzard2018, Grimm2020, Lescanne2020, Frattini2022, Berdou2023, Milul2023, Reglade2023, Marquet2023}.

Like other superconducting qubits based on Josephson junctions, Schrödinger cat qubits are subject to various noise sources, such as photon escape, dielectric loss, and, finally, residual Bogolyubov quasiparticles. Even though superconducting qubits are operated at very low temperatures, so hardly any quasiparticles should be present in thermal equilibrium, typically a significant number of residual non-equilibrium quasiparticles can still be detected. Presumably generated by rare energetic events (such as cosmic rays~\cite{Cardani2021}), dilute quasiparticles recombine very slowly, and it is well established that their density (normalized to the Cooper pair density) is usually in the range $x_\text{qp}\sim 10^{-5}-10^{-8}$ \cite{Martinis2009, Vool2014, Wang2014, Serniak2018}. Many experiments studying the coherence of transmon or fluxonium qubits \cite{Paik2011, Riste2013, Pop2014} are successfully described by taking into account residual quasiparticles via the theory developed in Refs.~\cite{Catelani2011PRL, Catelani2011, Catelani2012}. As qubits are improved by eliminating other error sources, Bogolyubov quasiparticles are likely to ultimately limit the coherence times. 

The fundamental difference between the conventional qubits, such as transmon or fluxonium, and cat qubits is that the former are based on stationary eigenstates of a static Hamiltonian, while the latter rely on a strong microwave drive. In fact, the qubit states are stationary only in a fast rotating reference frame, whose frequency is determined by a device-dependent combination of the natural frequencies of the circuit and the drive. The cat qubit states may be eigenstates of an engineered Kerr-like bosonic Hamiltonian~\cite{Grimm2020}, or form the stationary manifold of a two-photon dissipative Lindbladian~\cite{Leghtas2015, Lescanne2020}. This poses the question of how the driven (Kerr  qubit) or driven-dissipative (dissipative qubit) nature of the Schr\"odinger cat qubits affects their interaction with the residual quasiparticles. The present paper is dedicated to a theoretical investigation of this question.

In the following, we start from the quasiparticle tunneling Hamiltonian as in Refs.~\cite{Catelani2011, Catelani2012} and calculate the rates of various errors in Kerr and dissipative cat qubits. To identify these errors, it is convenient to use the phenomenological master equation, which became a standard tool for the description of the cat qubits' dynamics \cite{Puri2019, Grimm2020, Puri2020, Frattini2022, Gautier2022, Putterman2022, Chamberland2022, Gravina2023}:
\begin{equation}\label{eq:masterequation}
\frac{d\hat\rho}{dt}=\mathcal{L}_0\hat\rho
+ \kappa_-\mathcal{D}[\hat{a}]\hat\rho
+ \kappa_+\mathcal{D}[\hat{a}^\dagger]\hat\rho
+ \kappa_\phi\mathcal{D}[\hat{a}^\dagger\hat{a}]\hat\rho.
\end{equation}
Here $\hat\rho$ is the density matrix in the Hilbert space of a harmonic oscillator, which encodes the qubit, with raising and lowering operators $\hat{a}^\dagger,\hat{a}$. The density matrix is written in the rotating frame, in which the Lindbladian superoperator appearing on the right-hand side of Eq.~(\ref{eq:masterequation}) is stationary. Its main part $\mathcal{L}_0$ actually defines the qubit: it is purely Hamiltonian for the Kerr qubit, $\mathcal{L}_0\hat\rho\equiv-{i}[H_K,\hat\rho]$ with Kerr Hamiltonian $H_K=-K(\hat{a}^\dagger{}^2-\alpha^2)(\hat{a}^2-\alpha^2)$, or purely dissipative for the dissipative qubit,  $\mathcal{L}_0\hat\rho\equiv\kappa_2\mathcal{D}[\hat{a}^2-\alpha^2]\hat\rho$, with the dissipator $\mathcal{D}[\hat{O}]\hat\rho\equiv\hat{O}\hat\rho\hat{O}^\dagger-(\hat{O}^\dagger\hat{O}\hat\rho+\hat\rho\hat{O}^\dagger\hat{O})/2$ for any operator~$\hat{O}$, in both cases parametrized by a number $\alpha$ (which can be taken real and positive without loss of generality).
The qubit computational space is spanned by the two coherent states $|{\pm\alpha}\rangle\equiv{e}^{\pm\alpha(\hat{a}^\dagger-\hat{a})}|0\rangle$ (weakly non-orthogonal for $\alpha\gg1$, $\langle\alpha|{-\alpha}\rangle=e^{-2\alpha^2}$), or, equivalently, by their orthogonal linear combinations (Schr\"odinger cats) $|\mathcal{C}_\alpha^\pm\rangle\equiv(|{\alpha}\rangle\pm|{-\alpha}\rangle)/\sqrt{2(1\pm{e}^{-2\alpha^2})}$. 
The two states form  a degenerate eigenspace of the Kerr Hamiltonian $H_K$,
while the four corresponding operators $|\mathcal{C}_\alpha^\sigma\rangle\langle\mathcal{C}_\alpha^{\sigma'}|$ with $\sigma,\sigma'=\pm$ form a stationary subspace of the dissipator $\mathcal{D}[\hat{a}^2-\alpha^2]$.
The coefficient $K$ or $\kappa_2$ determines the intrinsic time scale of the Lindbladian~$\mathcal{L}_0$, which is set by the inverse of
either the energy gap $\omega_0\sim{K}\alpha^2$ separating the excited states of the Kerr Hamiltonian from the qubit subspace for the Kerr qubit, or the relaxation rate $\omega_0\sim\kappa_2\alpha^2$ towards the qubit subspace for the dissipative qubit.
The last three terms in Eq.~(\ref{eq:masterequation}) describe undesired relaxation processes for the qubit, characterized by the rates $\kappa_\mp$ (typically referred to as single-photon loss and gain rates) and $\kappa_\phi$ (pure dephasing rate), and lead to various errors.
Typically, the photon loss rate~$\kappa_-\gg\kappa_+,\kappa_\phi$.

A master equation similar to (\ref{eq:masterequation}) is often used for conventional static qubits (like transmon or fluxonium), written for the $2\times2$ qubit density matrix, where the harmonic oscillator operators have to be replaced by the Pauli operators as $\hat{a},\hat{a}^\dagger\to\hat\sigma_\mp\equiv(\hat\sigma_x\mp{i}\hat\sigma_y)/2$,  $\hat{a}^\dagger\hat{a}\to(\hat\sigma_z+1)/2$, $\mathcal{L}_0\hat\rho=-i(\omega_\text{qb}/2)[\hat\sigma_z,\hat\rho]$ with the qubit frequency $\omega_\text{qb}$ typically in the GHz range; the quasiparticle contribution to the corresponding rates was derived in Refs.~\cite{Catelani2011PRL, Catelani2011, Catelani2012}, where it was shown to be determined by the frequency-dependent normalized quasiparticle current spectral density $S_\text{qp}(\omega)$. Thus, it is natural to set the goal of calculating the coefficients $\kappa_\pm,\kappa_\phi$ in Eq.~(\ref{eq:masterequation}) due to quasiparticle tunneling, and to see how they differ from those in Refs.~\cite{Catelani2011PRL, Catelani2011, Catelani2012}. These coefficients uniquely determine the error rates of the qubit.

The standard derivation of the master equation gives Eqs. \eqref{eq:rates} for the rates. (There are important subtleties that we address below.) 
As in conventional qubits, the rates are determined by $S_\text{qp}(\omega)$, but the characteristic frequencies can be different. (i)~For the photon loss rate $\kappa_-$, we find that the relevant frequency is that of the rotating frame, while for static qubits it was the energy difference between the two qubit states. In practice, the two are of the same order of magnitude (in the GHz range, i.~e.,~the typical frequency scale for superconducting circuits), so $\kappa_-$ is quite similar for static and driven qubits. This is not surprising, since in both cases the error is due to the same physical process: a quasiparticle absorbs energy from the qubit while tunneling across a Josephson junction. (ii)~For the photon gain rate~$\kappa_+$ the situation is already different: while for static qubits the relevant frequency was negative (the quasiparticle had to give energy to the qubit, so the rate was strongly suppressed by the corresponding Boltzmann factor), for driven cat qubits we find a contribution at a positive frequency, which corresponds to a quasiparticle taking energy from the drive, and thus not subject to thermal suppression. (iii)~The coefficient $\kappa_\phi$, when calculated at the leading perturbative level, formally involves $S_\text{qp}(\omega=0)$ which is logarithmically divergent. For static qubits, Ref.~\cite{Catelani2012} proposed to cut off this divergence by $\kappa_\phi$ itself; subsequently, this argument was refined in Ref.~\cite{Zanker2015}, where resummation of an infinite subseries of the perturbation theory resulted in a non-exponential decay of the qubit coherence. For Schr\"odinger cat qubits, we find that the main effect of the dephasing term is captured if the logarithmic divergence is cut off at the intrinsic frequency scale $\omega_0$ of the qubit, as discussed above: $\omega_0\sim{K}\alpha^2$ for the Kerr qubit, or $\omega_0\sim\kappa_2\alpha^2$ for the dissipative qubit. The remaining terms have relative smallness $\sim\alpha^2{e}^{-2\alpha^2}$, for which the logarithmic divergence in $S_\text{qp}(\omega\to0)$ is not cut off by the intrinsic qubit dynamics, lead to the same problem as addressed in Refs.~\cite{Catelani2012,Zanker2015}.

Moreover, with the dissipator $\kappa_\phi\mathcal{D}[\hat{a}^\dagger\hat{a}]$, as written for the whole Hilbert space of the harmonic oscillator, Eq.~(\ref{eq:masterequation}) is not valid for the dissipative qubit, strictly speaking.
The reason is that the qubit-quasiparticle coupling has to be included perturbatively on top of the dissipative zeroth-order dynamics, in contrast to the usual situation when dissipative terms in the master  equation represent a perturbation with respect to a Hamiltonian dynamics.
As a result, the expression~(\ref{eq:kappa_phi}) for $\kappa_\phi$ only makes sense when Eq.~(\ref{eq:masterequation}) above is properly projected onto the qubit subspace.

Another point to stress is that Eq.~(\ref{eq:masterequation}), although being a convenient tool to study qubit errors, is not complete, formally speaking. 
Namely, one may, in principle, add higher-order dissipators $\mathcal{D}[\hat{a}^\dagger{}^m\hat{a}^n]$ with $n,m>1$, as well as Hamiltonian-type perturbations of the form $-i[\hat{h},\hat\rho]$, with $\hat{h}$ being a Hermitian combination of $\hat{a},\hat{a}^\dagger$, all of them inequivalent to each other when operating in the full Hilbert space of the oscillator. And indeed, below we find that the qubit coupling to quasiparticles generates a whole series of such terms. However, in the qubit subspace their effect reduces to small corrections with respect to the dissipators already appearing in Eq.~(\ref{eq:masterequation}), under the same assumptions that are used in the construction of the qubit itself, namely, the smallness of the superconducting phase fluctuations.

In the following section, we discuss these subtleties in detail, after defining the model and introducing the key quantities. In particular, we relate the eigenvalues of the Lindbladian of Eq.~(\ref{eq:masterequation}), which determine the qubit errors, with the coefficients $\kappa_\pm, \kappa_\phi$, see Eqs.~\eqref{eqs:lambdaDissip} and \eqref{eqs:lambdaKerr} for the dissipative and Kerr qubit, respectively. The derivation of the results is presented in Sec.~\ref{sec:DerivationKerr} for the Kerr qubit and in Sec.~\ref{sec:DerivationDissip} for the dissipative qubit, where we also compare our results to some recent experiments.
Finally, in Sec.~\ref{sec:Kinetics} we investigate the possibility of quasiparticle overheating by the drive, using the approach of Ref.~\cite{Catelani2019}. 
Even though the main interest of this paper is to derive Eq.~(\ref{eq:masterequation}) rather than to solve it, some properties of its solutions are relevant for the discussion, so we present some technical details regarding Eq.~(\ref{eq:masterequation}) in two appendices. 

\section{Model and summary of the main results}
\label{sec:Model}

Throughout the paper, we use units where the Planck and Boltzmann constants $\hbar=k_\text{B}=1$.

\subsection{Quasiparticle-qubit coupling}
\label{ssec:qubit-qp}

In this subsection, we show how the model of Refs.~\cite{Catelani2011PRL,Catelani2011} is adapted to Schr\"odinger cat qubits.

There are several different experimental realizations of such qubits \cite{Leghtas2015, Touzard2018, Grimm2020, Lescanne2020, Frattini2022, Berdou2023, Milul2023, Reglade2023, Marquet2023, Campagne-Ibarcq2020}. All of them contain one or several Josephson junctions connecting several superconducting islands. 
We label the islands by an index~$\iota$, the quasiparticle states on each island by $k$, and for brevity we omit the spin index whose role amounts to the usual factor of~2 in the rates. Assuming all islands $\iota$ to have the same superconducting gap~$\Delta$, and focusing on quasiparticle energies close to the gap, we write the quasiparticle Hamiltonian as
\begin{equation}\label{eq:Hqp}
\hat{H}_\mathrm{qp} = \sum_{\iota\in\{\text{islands}\}}\sum_k
\epsilon_{\iota,k}\hat\gamma^\dagger_{\iota,k}\hat\gamma_{\iota,k},\quad
\epsilon_{\iota,k}\approx\frac{\xi_{\iota,k}^2}{2\Delta},
\end{equation}
where $\hat\gamma^\dagger_{\iota,k}$ and $\hat\gamma_{\iota,k}$, respectively, are the creation and annihilation operators for quasiparticles with energy $\epsilon_{\iota,k}$ measured from $\Delta$, and $\xi_{\iota,k}$ are the electron energies measured from the Fermi level in the normal state. They determine the normal density of states $\nu_0$ per spin projection or, equivalently, the inverse mean level spacing~$\delta_\iota$, on each island,
\begin{equation}\label{eq:meanlevelspacing}
\frac{1}{\delta_\iota}=\sum_k\delta(\xi_{\iota,k}),
\end{equation}
assumed to be energy-independent and proportional to the island volume~$V_\iota$: $1/\delta_\iota=\nu_0V_\iota$. Here we assume $\nu_0$ to be the same for all islands.

As in Refs.~\cite{Catelani2011PRL,Catelani2011}, we assume the quasiparticle density $n_\text{qp}$ or, equivalently, the dimensionless concentration $x_\text{qp}\equiv{n}_\text{qp}/(2\nu_0\Delta)$ to be fixed by some external processes, and not by thermal equilibrium. The distribution of these quasiparticles over the energy levels $f(\epsilon_{\iota,k})$ will be assumed to be determined by the phonon temperature $T\sim10-30\:\mbox{mK}$ \cite{Leghtas2015, Touzard2018, Grimm2020, Lescanne2020, Frattini2022, Berdou2023, Milul2023, Reglade2023, Marquet2023} in most of the paper (except in Sec.~\ref{sec:Kinetics} where we study possible deviations from the thermal distribution due to the drive). Then the occupation probability $f(\epsilon_{\iota,k})$ of each energy level is $f(\epsilon_{\iota,k})=f_T(\epsilon_{\iota,k})$ with
\begin{equation}\label{eq:fthermal}
f_T(\epsilon)=x_\text{qp}\sqrt{\frac{\Delta}{2\pi{T}}}\,e^{-\epsilon/T},
\end{equation}
so that the density $2\nu_0\int_{-\infty}^\infty{d\xi}\,f_T(\xi^2/2\Delta)$ matches the given~$n_\text{qp}$.

The quasiparticles couple to the superconducting degrees of freedom when they tunnel across the Josephson junctions. Labeling the junctions by~$j$ and denoting the superconducting phase difference across each junction~$j$ by $\hat\varphi_j$, we have the coupling Hamiltonian~\cite{Catelani2011PRL,Catelani2011}:
\begin{align}
\hat{H}_\mathrm{Jqp} = {}&{}\sum_{j\in\{\text{junctions}\}}\sum_{k,k'}
\mathcal{T}_{j,kk'}\hat\gamma^\dagger_{\iota_{jL},k}\hat\gamma_{\iota_{jR},k'}\nonumber{}\\
{}&{}\quad\times\left(u_{\iota_{jL},k}u_{\iota_{jR},k'}e^{i\varphi_j/2}
-v_{\iota_{jL},k}v_{\iota_{jR},k'}e^{-i\varphi_j/2}\right)\nonumber{}\\
{}&{}+\mbox{H.\:c.}.
\label{eq:HJqp}
\end{align}
Here $\iota_{jL}$ and $\iota_{jR}$ are the two islands forming junction~$j$, and $u_{\iota,k},v_{\iota,k}$ are the Bogolyubov coefficients:
\begin{equation}
v_{\iota,k}^2=1-u_{\iota,k}^2
=\frac{1}2-\frac12\,\frac{\xi_{\iota,k}}{\sqrt{\Delta^2+\xi_{\iota,k}^2}}.
\end{equation}
The tunneling matrix elements $\mathcal{T}_{j,kk'}$, assumed to be real and energy-independent (on the relevant scale $\Delta$), determine the Josephson energy $E_{Jj}$ of the corresponding junction by the Ambegaokar-Baratoff relation~\cite{Ambegaokar1963}:
\begin{equation}
\sum_{k,k'}\mathcal{T}_{j,kk'}^2
\delta(\xi_{k'})\,\delta(\xi_k)
=\frac{E_{Jj}}{\pi^2\Delta}.
\end{equation}
If we focus on low energies, $|\xi_{\iota,k}|\ll\Delta$, then $u_{\iota,k}\approx{v}_{\iota,k}\approx1/\sqrt{2}$, and the Hamiltonian assumes a simpler form:
\begin{subequations}\begin{align}
\hat{H}_\mathrm{Jqp} &= \sum_{j\in\{\text{junctions}\}}\hat{\mathcal{I}}_j\sin\frac{\hat\varphi_j}2,
\label{eq:perturbationsin}\\
\hat{\mathcal{I}}_j&\equiv
i\sum_{k,k'}\mathcal{T}_{j,kk'}
\left(\hat\gamma^\dagger_{\iota_{jL},k}\hat\gamma_{\iota_{jR},k'}
-\hat\gamma^\dagger_{\iota_{jR},k'}\hat\gamma_{\iota_{jL},k}\right).
\label{eq:Ij}
\end{align}\end{subequations}
The operator $\hat{\mathcal{I}}_j$ is the quasiparticle contribution to the electric current through the junction (up to a factor of the electron charge).
As in Refs.~\cite{Catelani2011PRL,Catelani2011}, the results will be expressed in terms of the normalized quasiparticle current spectral density in each junction,
\begin{equation}\label{eq:Sqpdef}
S_{\text{qp},j}(\omega)\equiv\int_{-\infty}^\infty\langle\hat{\cal I}_j(t)\,\hat{\cal I}_j(0)\rangle\,e^{i\omega{t}}\,dt,
\end{equation}
where the time dependence is determined by the Hamiltonian $\hat{H}_\mathrm{qp}$ in Eq.~(\ref{eq:Hqp}):
$\hat{\cal I}_j(t)\equiv{e}^{i\hat{H}_\mathrm{qp}t}\hat{\cal I}_j{e}^{-i\hat{H}_\mathrm{qp}t}$.
This spectral density indicates the probability to absorb a quantum of energy~$\omega$ at the junction~$j$.
For the thermal distribution~(\ref{eq:fthermal}) and to the leading order in $|\omega|/\Delta$, $T/\Delta$, $S_{\text{qp},j}(\omega)$ evaluates to
\begin{equation}\label{eq:Sthermal}
S_{\text{qp},j}(\omega)=x_\text{qp}\,\frac{16E_{Jj}}\pi
\sqrt{\frac{\Delta}{2\pi{T}}}\,e^{\omega/2T}\,K_0\!\left(\frac{|\omega|}{2T}\right),
\end{equation}
where $K_0(z)$ is the modified Bessel function. At large positive $\omega\gg{T}$ this expression is slowly decaying, $e^{z}\,K_0(|z|)\sim\sqrt{\pi/(2|z|)}$, while at large negative frequencies it is exponentially suppressed, $e^{z}\,K_0(|z|)\sim\sqrt{\pi/(2|z|)}\,e^{-2|z|}$, since the quasiparticle has to emit energy.
At small $|\omega|\ll{T}$, $S_{\text{qp},j}(\omega)$ is logarithmically divergent: $e^{z}\,K_0(|z|)\sim\ln(1/|z|)$.
 
The qubit degree of freedom is represented by a combination of phases $\hat\varphi_j$ on several junctions, which depends on the specific device architecture. In the fast rotating frame the corresponding dynamical variable is convenient to express in terms of the harmonic oscillator raising and lowering operators $\hat{a}^\dagger,\hat{a}$, which appear in the master equation \eqref{eq:masterequation} and whose dynamics is slow. Then, each phase $\hat\varphi_j$ can be represented as
\begin{align}
\hat\varphi_j={}&{}\varphi_{\text{bias},j}
+\varphi_{\text{d},j}e^{-i\omega_\text{d}t}+\varphi_{\text{d},j}^*e^{i\omega_\text{d}t}\nonumber\\
{}&{}+\left(\varphi_{a,j}\hat{a}e^{-i\omega_at} + \varphi_{a,j}^*\hat{a}^\dagger{e}^{i\omega_at}\right)+\ldots.
\label{eq:phij}
\end{align}
Here $\varphi_{\text{bias},j}$ is a constant phase bias, not necessarily small, controlled by an external flux. $\omega_\text{d}$ and $\varphi_{\text{d},j}$ are the frequency of the external classical drive and its amplitude on the $j$th junction. $\varphi_{a,j}$  is the amplitude of the qubit mode on the $j$th junction, and $\omega_a$ is the frequency of the rotating frame. This frequency is determined by the requirement that the dynamics of $\hat{a}^\dagger,\hat{a}$ is slow, and is given by a device-dependent combination of the natural frequencies of the circuit and the drive. Finally, ``$\ldots$'' stands for terms involving other degrees of freedom of the circuit, which are orthogonal to the qubit mode; they are weakly coupled and strongly detuned in energy (on the scale of the qubit dynamics), so their effect can be neglected.

We make the crucial assumption that $|\varphi_{a,j}| \ll 1/\alpha$ and $|\varphi_{\text{d},j}|\ll 1$, to expand $\sin(\hat\varphi_j/2)$ in Eq.~\eqref{eq:perturbationsin}; this holds when all junctions are sufficiently large, such that their Josephson energy exceeds the charging energy.
We note that the same assumption underlies the construction of the cat qubits themselves~\cite{Leghtas2013, Leghtas2015, Mirrahimi2014, Touzard2018, Grimm2020, Lescanne2020, Frattini2022, Berdou2023, Milul2023, Reglade2023, Marquet2023}. Indeed, the zeroth-order Lindbladian~$\mathcal{L}_0$ is obtained using the expansion of the Josephson nonlinearity to several low orders.

\subsection{Error rates in the phenomenological master equation}
\label{ssec:phenomenological}

The phenomenological master equation~(\ref{eq:masterequation}) includes error dissipators that can be naturally related to various physical mechanisms, such as photon escape to an external circuit for single-photon loss $\kappa_-\mathcal{D}[\hat{a}]$, high-energy photons due to poor filtering for single-photon gain $\kappa_+\mathcal{D}[\hat{a}^\dagger]$, coupling to two-level systems for pure dephasing $\kappa_\phi\mathcal{D}[\hat{a}^\dagger\hat{a}]$). The main subject of this paper is the microscopic derivation of the quasiparticle contribution to the different error dissipators. 
In this subsection, we briefly discuss qubit errors due to these dissipators, as found by the approximate solution of Eq.~(\ref{eq:masterequation}) within the computational subspace. The error rates are obtained by assuming $\kappa_+,\kappa_-,\kappa_\phi$ to be small and treating the corresponding terms as perturbations on top of the main term~$\mathcal{L}_0$.
Most of these results are known~\cite{Puri2019, Grimm2020, Puri2020, Frattini2022, Gautier2022, Putterman2022, Chamberland2022, Gravina2023} (see also Ref.~\cite{LeRegent2024} for a systematic perturbation expansion, implemented numerically).

To recall the general structure of the degenerate Lindbladian perturbation theory \cite{Stenholm1986, Cirac1992, Reiter2012, Kessler2012, Cai2013, Medvedyeva2016, LeRegent2024}, if the master equation is of the form
$\partial\hat\rho/\partial{t}=\mathcal{L}_0\hat\rho+\mathcal{L}_1\hat\rho$
with $\mathcal{L}_1\ll\mathcal{L}_0$, then we can define two subspaces $S_\|$ and $S_\perp$, such that $\mathcal{L}_0\hat\rho=0$ for all $\hat\rho\in{S}_\|$, and the equation $\mathcal{L}_0\hat{x}=\hat\rho$ with unknown $\hat{x}$ has solutions for all $\hat\rho\in{S}_\perp$. Equivalently, $S_\|$ and $S_\perp$ are spanned by two sets of right eigenvectors with zero and non-zero eigenvalues, respectively. Then, any density matrix can be split as $\hat\rho=\hat\rho_\|+\hat\rho_\perp$, which defines the projectors $\mathcal{P}_\|$ and $\mathcal{P}_\perp = 1 - \mathcal{P}_\|$, such that $\mathcal{L}_0\mathcal{P}_\|=\mathcal{P}_\|\mathcal{L}_0=0$. (Note that $\mathcal{P}_\|$ can be viewed as the result of the evolution $e^{\mathcal{L}_0t}$ at $t\to\infty$ for a dissipative $\mathcal{L}_0$.) The perturbation $\mathcal{L}_1$ induces nontrivial dynamics in the slow subspace. This slow subspace is a weakly deformed $S_\|$, such that the density matrix has a small component in $S_\perp$, namely, 
\begin{equation}\label{eq:rhoperp=}
\hat\rho_\perp=-\mathcal{P}_\perp\mathcal{L}_0^{-1}\mathcal{P}_\perp\mathcal{L}_1\hat\rho_\|+O(\mathcal{L}_1^2),
\end{equation}
and the dynamics is determined by the projected master equation
\begin{equation}\label{eq:masterprojected}
\frac{\partial\hat\rho_\|}{\partial{t}}=\mathcal{P}_\|\mathcal{L}_1\hat\rho_\|
-\mathcal{P}_\|\mathcal{L}_1\mathcal{P}_\perp\mathcal{L}_0^{-1}\mathcal{P}_\perp\mathcal{L}_1\hat\rho_\| + O(\mathcal{L}_1^3).
\end{equation}
The component $\hat\rho_\perp$ then follows adiabatically according to Eq.~(\ref{eq:rhoperp=}). It determines the small but finite probability to find the system outside $S_\|$ at any instant of time.

 For the dissipative qubit, the zero subspace $S_\|$ of $\mathcal{L}_0$  is spanned by four matrices $|\mathcal{C}_\alpha^\sigma\rangle\langle\mathcal{C}_\alpha^{\sigma'}|$ with $\sigma,\sigma'=\pm$. The first-order term in the projected master equation can be found using the known left eigenvectors of $\mathcal{L}_0$ corresponding to the zero eigenvalue (see Ref.~\cite{Guillaud2023} and Appendix~\ref{app:perturbations}).
Taking into account the perturbation $\mathcal{L}_1$, three out of four eigenvalues become non-zero; their large-$\alpha$ asymptotes are 
\begin{subequations}\label{eqs:lambdaDissip}\begin{align}
\lambda_x = {}&{} -2\kappa_-\alpha^2 - 2\kappa_+(\alpha^2+1),\\
\lambda_y = {}&{} -2\kappa_-\alpha^2 - 2\kappa_+(\alpha^2+1)
             -2\kappa_\phi\alpha^2e^{-2\alpha^2},\\
\lambda_z = {}&{} -2\kappa_-\alpha^2e^{-4\alpha^2}
             -2\kappa_+e^{-2\alpha^2}
             -2\kappa_\phi\alpha^2e^{-2\alpha^2}\nonumber\\
             {}&{} -\frac{\kappa_-^2}{\kappa_2}\,e^{-2\alpha^2}.
\end{align}\end{subequations}
The eigenvalue $\lambda_z$ corresponds to the eigenvector $|\mathcal{C}_\alpha^+\rangle\langle\mathcal{C}_\alpha^-|+|\mathcal{C}_\alpha^-\rangle\langle\mathcal{C}_\alpha^+|
\propto|\alpha\rangle\langle\alpha|-|{-\alpha}\rangle\langle-\alpha|$. Its exponential smallness, in contrast with $\lambda_x$ and $\lambda_y$, is a manifestation of the suppressed probability to transfer population between $|\alpha\rangle$ and $|{-\alpha}\rangle$ by a local perturbation, because of their small overlap. This strong asymmetry between different rates is a general feature of cat qubits, which allows for an efficient implementation of quantum error correction codes \cite{Guillaud2023}. 
The suppression is especially strong for the first-order photon loss, $\propto{e}^{-4\alpha^2}$; however, in the second order in $\kappa_-$ the standard factor $e^{-2\alpha^2}$ is restored \cite{Dubovitskii2024}, as found by evaluating the second term in Eq.~\eqref{eq:masterprojected}.


The leakage probability,
\begin{equation}\label{eq:werr}
w_\text{leak}\equiv 1- \sum_\sigma\langle\mathcal{C}_\alpha^\sigma|\hat\rho|\mathcal{C}_\alpha^\sigma\rangle,
\end{equation}
for the dissipative qubit is determined by $\hat\rho_\perp$ from Eq.~(\ref{eq:rhoperp=}) as $w_\text{leak}=-\sum_\sigma\langle\mathcal{C}_\alpha^\sigma|\hat\rho_\perp|\mathcal{C}_\alpha^\sigma\rangle$.
Indeed, since $\Tr\hat\rho=1$ is conserved, $\Tr\hat\rho_\perp=0$, and thus $\sum_\sigma\langle\mathcal{C}_\alpha^\sigma|\hat\rho_\||\mathcal{C}_\alpha^\sigma\rangle=1$. The photon loss $\mathcal{D}[\hat{a}]$ does not produce any leakage in the first order, while $\mathcal{D}[\hat{a}^\dagger]$ and $\mathcal{D}[\hat{a}^\dagger\hat{a}]$ yield a finite leakage probability, $\sim\kappa_+/\kappa_2$ and $\sim\alpha^2\kappa_\phi/\kappa_2$, respectively.

For the Kerr qubit the zero subspace of $\mathcal{L}_0$ is much larger: in addition to the four-dimensional qubit subspace, it includes all matrices that are diagonal in the basis of the eigenvectors of the Kerr Hamiltonian $H_K$.
Moreover, at large $\alpha$ low eigenstates come in doublets with exponentially small energy splitting; this makes the off-diagonal matrix elements within each doublet also slow. Then, for $\hat\rho$ in the computational space spanned by $|\mathcal{C}_\alpha^\sigma\rangle\langle\mathcal{C}_\alpha^{\sigma'}|$, the projection $\mathcal{P}_\|\,\mathcal{D}[\hat{a}]\hat\rho$ is also in the computational space, while $\mathcal{P}_\|\,\mathcal{D}[\hat{a}^\dagger]\hat\rho$ and $\mathcal{P}_\|\,\mathcal{D}[\hat{a}^\dagger\hat{a}]\hat\rho$ have components on other states as well, which corresponds to probability leakage outside the computational space. As a result, if one simply projects $\mathcal{L}_1$ on the computational subspace by brute force, 
(i)~there is no exponential suppression of the $\kappa_+,\kappa_\phi$ contribution to the eigenvalue~$\lambda_z$, and (ii)~all four eigenvalues are non-zero, the finite value of $\lambda_0$ representing the leakage rate:
\begin{subequations}\label{eqs:lambdaKerr}
\begin{align}
&\lambda_x = -2\kappa_-\alpha^2 - \kappa_+(2\alpha^2+1) - \kappa_\phi\alpha^2,\\
&\lambda_y = -2\kappa_-\alpha^2 - \kappa_+(2\alpha^2+1)
             -\kappa_\phi\alpha^2,\\
&\lambda_z = -2\kappa_-\alpha^2e^{-4\alpha^2}
             -\kappa_+
             -\kappa_\phi\alpha^2,\label{eq:lambda_xKerrProjected}\\
&\lambda_0 = -\kappa_+
             -\kappa_\phi\alpha^2.
\end{align}\end{subequations}
Since typically $\kappa_-\gg\kappa_+,\kappa_\phi$, the eigenvalue $\lambda_z$ may still be dominated by the first term; at the same time, the finite leakage rate $-\lambda_0$ implies that all probability would eventually leak out of the computational subspace.
However, we should recall that Eqs.~(\ref{eqs:lambdaKerr}) do not represent the true error rates, since the leaked probability is brought back to the computational subspace by the one-photon loss $\mathcal{D}[\hat{a}]$, and this process may occur without error.
As a result, the stationary state of the full Lindbladian contains a small probability $\sim\kappa_+/\kappa_-,\kappa_\phi/\kappa_-$ outside the computational space. Numerics shows that the first non-zero eigenvalue decreases with growing $\alpha$ in a step-like fashion, with an exponential envelope, $\sim{e}^{-C\alpha^2}$ with $C\approx0.8$~\cite{Gautier2022, Frattini2022}. Thus, the true error rates can be much smaller than predicted by Eqs.~(\ref{eqs:lambdaKerr}).

\subsection{Quasiparticle-induced error rates}
\label{ssec:quasiparticle-induced}

Assuming no correlation between the quasiparticle currents $\hat{\cal I}_j$ on different junctions, we find that their contributions to various error rates add up incoherently. Thus, in the following we will omit the junction index~$j$ everywhere, as if there were only one Josephson junction in the system. If there are several junctions, one should restore the index~$j$ and sum the corresponding rates over~$j$.

The standard perturbative derivation of the master equation (see, e.~g., Ref.~\cite{BreuerPetruccione}) results in the coefficient at a dissipator $\mathcal{D}[\hat{O}]$ such that the master equation reproduces the rates of the bath-induced transitions between energy levels of the unperturbed system, as given by Fermi's golden rule with the perturbation~$\hat{O}$. Since the master equation~(\ref{eq:masterequation}) is written in the rotating frame, we have to substitute $\hat\varphi$ from Eq.~(\ref{eq:phij}) with fast oscillating terms in Eq.~(\ref{eq:perturbationsin}), and apply the golden rule for periodic perturbations to different terms in the expansion of $\sin(\hat\varphi/2)$ oscillating at different frequencies. In principle, this procedure leads to an infinite series of dissipators of the form $\mathcal{D}[\hat{a}^\dagger{}^n\hat{a}^m]$.

Under the assumption $|\varphi_a|\alpha\ll1$, high-order dissipators are weak. Among the first-order contributions, the strongest one is the photon loss $\mathcal{D}[\hat{a}]$. The photon gain $\mathcal{D}[\hat{a}^\dagger]$, although weaker, leads to a qualitatively different effect for the Kerr qubit, namely, leakage out of the computational space, as discussed in the previous subsection. 
In the first order, coupling to quasiparticles also produces a Hamiltonian correction $\propto{i}[\hat{a}^\dagger\hat{a},\hat\rho]$; however, it can be removed by adjusting the rotating frame frequency.
Second-order contributions, although weaker, have a different symmetry: while single-photon loss (gain) switches the photon parity $(-1)^{\hat{a}^\dagger\hat{a}}$, the second-order terms preserve parity, and thus may require a different error correction scheme.
Among these, the pure dephasing $\mathcal{D}[\hat{a}^\dagger\hat{a}]$ is the most important; indeed, the two-photon loss, $\mathcal{D}[\hat{a}^2]$, acts trivially in the qubit subspace, while the two-photon absorption, $\mathcal{D}[\hat{a}^\dagger{}^2]$, is weak due to the lack of high-energy quasiparticles (by the same smallness as in the single-photon processes, $\kappa_+/\kappa_-\ll1$).
These are the three processes included in Eq.~(\ref{eq:masterequation}), and our result for the coefficients is
\begin{subequations} \label{eq:rates}
\begin{align}
&\kappa_- = S_\text{qp}(\omega_a)\,\frac{|\varphi_a|^2}4\cos^2\frac{\varphi_\text{bias}}{2},
\label{eq:kappa_minus}\\
&\kappa_+ = S_\text{qp}(\omega_\text{d}-\omega_a)\,\frac{|\varphi_a|^2}4\,\frac{|\varphi_\text{d}|^2}4\sin^2\frac{\varphi_\text{bias}}{2},
\label{eq:kappa_plus}\\
&\kappa_\phi = S_\text{qp}(\omega_0)\,\frac{|\varphi_a|^4}{16}\sin^2\frac{\varphi_\text{bias}}{2},
\label{eq:kappa_phi}
\end{align}
\end{subequations} 
calculated to the leading order in tunneling and $\omega/\Delta$, and valid
up to some details to be discussed in the following paragraphs. 

The $\varphi_\text{bias}$ dependence in Eqs. \eqref{eq:rates} reflects quasiparticle interference in the error process, like in static qubits~\cite{Pop2014}. In cat qubits, $\varphi_\text{bias}$ determines the working point and cannot be easily adjusted.
Typically, $\varphi_\text{bias}$ is some generic number of the order of unity, but in some devices it may be close to $0$ or $\pi$, so some rates vanish. Then one has to include corrections to Eqs.~(\ref{eq:perturbationsin}) and (\ref{eq:Sthermal}), subleading in $\xi_k/\Delta$ and $|\omega|/\Delta$, respectively~\cite{Catelani2012, Catelani2014}, that would make the rates finite.

Most of the terms in the expansion of $\sin(\hat\varphi/2)$ oscillate at frequencies far exceeding~$\omega_0$, the frequency scale of $\mathcal{L}_0$; then, in the first approximation one can neglect $\mathcal{L}_0$ and write the golden rule as if the mode~$\hat{a}$ corresponded to a harmonic oscillator with zero frequency, and all energy were taken or given by the quasiparticle. Equation~(\ref{eq:kappa_minus}) is obtained by picking the leading term $(\varphi_a/2)\hat{a}e^{-i\omega_at}\cos(\varphi_\text{bias}/2)$ in the expansion of $\sin(\hat\varphi/2)$; its effect is obviously described by the photon loss dissipator $\kappa_-\mathcal{D}[\hat{a}]$ since the rates of all transitions induced by this perturbation are proportional to the same factor $S_\text{qp}(\omega_a)$, independent of the  transition. 

By analogy, the photon gain dissipator $\kappa_+\mathcal{D}[\hat{a}^\dagger]$ can be obtained from the conjugate term $(\varphi_a^*/2)\hat{a}^\dagger{e}^{i\omega_at}\cos(\varphi_\text{bias}/2)$. This results in an expression for $\kappa_+$ obtained from Eq.~(\ref{eq:kappa_minus}) by the replacement $S_\text{qp}(\omega_a)\to{S}_\text{qp}(-\omega_a)$. This corresponds to a quasiparticle emitting energy $\omega_a$, whose rate is proportional to the population of such high-energy quasiparticles. For quasiparticles in equilibrium with phonons at temperature~$T$, it is proportional to $e^{-\omega_a/T}$, a rather small factor for typical parameters ($\omega_a \sim $ a few GHz, $T\sim 10 - 30$ mK).
We find a large contribution to $\mathcal{D}[\hat{a}^\dagger]$ originating from a higher-order term in the expansion of $\sin(\hat\varphi/2)$:  $-(\varphi_a^*/2)\hat{a}^\dagger{e}^{i\omega_at}\,(\varphi_{\text{d}}/2)e^{-i\omega_\text{d}t}\,\sin(\varphi_\text{bias}/2)$. 
Typically, the drive frequency $\omega_\text{d}>\omega_a$, so this term oscillates at a positive frequency; during a single tunneling process a drive photon is absorbed by the quasiparticle which then immediately emits a qubit resonator photon.\footnote{
If $\omega_\text{d}<\omega_a$, the same absence of the Boltzmann factor is found in a higher order $(p+1)$, such that $p\omega_\text{d}>\omega_a$.
} 
For typical experimental parameters, the smallness introduced by going to the next order in the expansion turns out to be well compensated by the absence of the factor $e^{-\omega_a/T}$. In the specific case $\omega_\text{d}=2\omega_a$ (relevant for the Kerr qubit), the perturbation oscillates at the same frequency as the photon loss. Then, instead of an incoherent sum of two dissipators, $\kappa_-\mathcal{D}[\hat{a}]+\kappa_+\mathcal{D}[\hat{a}^\dagger]$, one has to mix the two terms coherently in a single dissipator $\mathcal{D}[\sqrt{\kappa_-}\,\hat{a}+\sqrt{\kappa_+}\hat{a}^\dagger]$, which, however, leads to practically the same results for errors, as we numerically check in Appendix~\ref{app:numerics}.

Generally, the assumption of quasiparticle equilibrium should not be taken for granted in a driven system. The quasiparticle distribution may be altered by absorption of drive photons, which competes with phonon emission and quasiparticle escape. The resulting non-thermal quasiparticle population at high energies can give another contribution to $\kappa_+$. The rates for photon absorption, phonon emission, and quasiparticle escape are quite dependent on the specific qubit architecture. In Sec.~\ref{sec:Kinetics}, we study this competition for qubits produced in Refs.~\cite{Grimm2020, Lescanne2020} and find the corresponding contribution to $\kappa_+$ to be smaller than the higher-order drive contribution discussed just above. However, the difference is less than an order of magnitude, so for some structures the non-thermal population may dominate the rate~$\kappa_+$.

The dephasing dissipator $\kappa_\phi\mathcal{D}[\hat{a}^\dagger\hat{a}]$ can be obtained in the leading order by expanding $\sin(\hat\varphi/2)$ to the second order and taking the cross-term $-(|\varphi_a|^2/4)\hat{a}^\dagger\hat{a}\sin(\varphi_\text{bias}/2)$. In contrast to the previous ones, this term does not oscillate, leading to a divergent quantity $S_\text{qp}(\omega=0)$. Then, to describe its effect, it is necessary to consider the unperturbed dynamics of the system, which is quite different for the Kerr and dissipative qubits.

For the Kerr qubit, whose unperturbed dynamics is Hamiltonian, the situation is rather standard. Namely, $\hat{a}^\dagger\hat{a}$ should be decomposed into components corresponding to transitions between different energy levels $\varepsilon_{l\sigma}$ of the Kerr Hamiltonian $H_K$. 
These levels are classified by their parity $\sigma=(-1)^{\hat{a}^\dagger\hat{a}}$, which is conserved by $H_K$, and an integer $l\geq0$, such that $\varepsilon_{0+}=\varepsilon_{0-}=0$ correspond to $|\mathcal{C}_\alpha^\pm\rangle$. Since the perturbation $\hat{a}^\dagger\hat{a}$ conserves parity, instead of the single dissipator $\mathcal{D}[\hat{a}^\dagger\hat{a}]$, the master equation contains a sum of dissipators, each one corresponding to a given transition, with the coefficient proportional to $S_\text{qp}(\varepsilon_{l\sigma}-\varepsilon_{l'\sigma})$. Since relevant levels are those with not too high $l,l'$, we can write $|\varepsilon_{l\sigma}-\varepsilon_{l'\sigma}|\sim|\varepsilon_{l\sigma}|\sim|\varepsilon_{l'\sigma}|\sim{K}\alpha^2\ll{T}$ for typical experimental parameters. Then, with logaritmic precision, we can write
\begin{equation}\label{eq:logprecision}
S_\text{qp}(\varepsilon_{l\sigma}-\varepsilon_{l'\sigma})\approx
x_\text{qp}\,\frac{16E_{J}}\pi\sqrt{\frac{\Delta}{2\pi{T}}}\ln\frac{T}{K\alpha^2},
\end{equation}
and wrap all components back into the dissipator $\mathcal{D}[\hat{a}^\dagger\hat{a}]$ whose coefficient is given by Eq.~(\ref{eq:kappa_phi}) with $\omega_0\sim{K}\alpha^2$. This procedure works for all transitions for which $l>0$ or $l'>0$; these transitions determine the leakage out of the computational space, which is the main effect of $\mathcal{D}[\hat{a}^\dagger\hat{a}]$ in the Kerr qubit, as discussed in the previous subsection and represented by the $-\kappa_\phi\alpha^2$ terms in Eqs.~(\ref{eqs:lambdaKerr}). Still, $\mathcal{D}[\hat{a}^\dagger\hat{a}]$ has an exponentially small component which acts directly in the computational space with $l=l'=0$; this component can be represented as $2\kappa_\phi\alpha^2e^{-2\alpha^2}\mathcal{D}[\sigma_z]$, where $\sigma_z$ is the Pauli matrix in the basis $|\mathcal{C}_\alpha^+\rangle,|\mathcal{C}_\alpha^-\rangle$, and $\kappa_\phi$ is proportional to $S_\text{qp}(0)$. For this component, the logarithmic divergence is not cut off by the Kerr Hamiltonian, and one has to invoke the same arguments as in Refs.~\cite{Catelani2012,Zanker2015} for stationary qubits. This issue, however, seems to be of little practical relevance for the cat qubits due to the exponential smallness of the divergent component and the weakness of the logarithmic divergence.



For the dissipative qubit, to derive the dephasing dissipator $\kappa_\phi\mathcal{D}[\hat{a}^\dagger\hat{a}]$, we face an unconventional task of developing perturbation theory in system-bath coupling on top of the dissipative unperturbed system dynamics. 
%
%
%
%
We find that the Markovian master equation~(\ref{eq:masterequation}), valid in the whole Hilbert space of the $\hat{a}$ oscillator, cannot be derived. We can only justify the first-order term in the projected master equation~(\ref{eq:masterprojected}), but not the second-order term. Still, from the practical point of view, the first-order term is sufficient as long as quasiparticles are dilute.

Physically, the first-order term in Eq.~(\ref{eq:masterprojected}) describes the following process: the qubit residing in the steady computational subspace $S_\|$ is suddenly hit by a quasiparticle, performing a transition to the orthogonal subspace~$S_\perp$, followed by the relaxation back to~$S_\|$ (we remind that the projector $\mathcal{P}_\|=\lim_{t\to\infty}e^{\mathcal{L}_0t}$). Since the levels in $S_\perp$ are broadened by the relaxation, the quasiparticle energy slightly changes during this process; namely, some energy can be taken from the drive and emitted into the bath which is responsible for the dissipative~$\mathcal{L}_0$. It is this inelastic process that smears the singularity in $S_\text{qp}(\omega\to0)$.

The second-order term in Eq.~(\ref{eq:masterprojected}) would correspond to a process when a second quasiparticle arrives quickly after the first one, so that the qubit has not yet relaxed back to~$S_\|$, and the two quasiparticles exchange energy with the drive and the bath independently from each other. However, in such process the quasiparticles may also exchange energy among themselves; to account for this, one has to go back to the original Hamiltonian and study the two-quasiparticle process in full detail. This would produce a contribution to the projected master equation which is of the same order, but does not reduce to the second-order term in Eq.~(\ref{eq:masterprojected}). In practice,  $x_\text{qp}$ is rather small, so such two-quasiparticle processes are too rare to be included into the scope of the present paper.

The derivation presented in Sec.~\ref{sec:DerivationDissip} results in an effective Lindbladian superoperator acting on $2\times2$ density matrices $\hat\rho_\|$ in the qubit coding subspace, which can be interpreted as the projection $\mathcal{P}_\|\kappa_\phi\mathcal{D}[\hat{a}^\dagger\hat{a}]\hat\rho_\|$, with logarithmic precision. Beyond the logarithmic precision, the dissipator should be decomposed into a sum of terms corresponding to eigenvalues $\lambda_m$ and eigenvectors of the dissipative Lindbladian~$\mathcal{L}_0$. Each such term is proportional to $\Re{S}_\text{qp}(i\lambda_m)$, instead of $S_\text{qp}(\varepsilon_{l\sigma}-\varepsilon_{l'\sigma})$ for the Kerr qubit. As for the Kerr qubit, there is also an exponentially small component which remains proportional to the logarithmically divergent $S_\text{qp}(0)$, to be handled as in Refs.~\cite{Catelani2012,Zanker2015} for stationary qubits.

\section{Dissipators for the Kerr qubit}
\label{sec:DerivationKerr}

In this section, we provide more details on the derivation of the different dissipators in the  case of the Kerr qubit. The derivation of the master equation follows the standard textbook route (see, e.~g., Ref.~\cite{BreuerPetruccione}). We start with the Hamiltonian of the decoupled system (the $\hat{a}$ mode) and bath (the quasiparticles),
\begin{equation}\label{eq:H0Kqp}
\hat{H}_0 \equiv\hat{H}_K + \hat{H}_\text{qp} = -K(\hat{a}^\dagger{}^2-\alpha^2)(\hat{a}^2-\alpha^2)
+\sum_{\iota,k}\epsilon_{\iota,k}\hat\gamma^\dagger_{\iota,k}\hat\gamma_{\iota,k},
\end{equation}
where $\iota=L,R$ denotes the two islands forming the Josephson junction (see the discussion in the end of Sec.~\ref{ssec:qubit-qp}), and the qubit-quasiparticle coupling perturbation in the form
\begin{equation}\label{eq:H1perturbation}
\hat{H}_1(t) = \hat{\cal I}\otimes
\left(\hat{A} e^{-i\Omega{t}}+\hat{A}^\dagger e^{i\Omega{t}}\right),
\end{equation}
where the current operator $\hat{\mathcal{I}}$ is given by Eq.~(\ref{eq:Ij}), and the factor in the brackets represents one of the terms of the expansion of $\sin(\hat\varphi/2)$, oscillating at a given frequency $\Omega>0$ determined by the corresponding combination of terms in Eq.~(\ref{eq:phij}). For example, we have a perturbation $\Omega=\omega_a$ and 
\begin{subequations}
\begin{align}
\hat{A}={}&{}\cos\frac{\varphi_\text{bias}}{2} \nonumber\\
{}&{}\times\left[\frac{\varphi_a}{2}\,\hat{a}-\frac{|\varphi_a|^2\varphi_a}{48}\left(\hat{a}\hat{a}\hat{a}^\dagger+\hat{a}\hat{a}^\dagger\hat{a}+\hat{a}^\dagger\hat{a}\hat{a}\right)+\ldots\right],\label{eq:A=a}
\end{align}
where we only keep the first term by virtue of the assumption $|\varphi_a|\alpha\ll1$. Another perturbation has $\Omega=\omega_\text{d}-\omega_a$ and
\begin{equation}\label{eq:A=adag}
\hat{A}=-\sin\frac{\varphi_\text{bias}}{2}\left(\frac{\varphi_a^*\varphi_\text{d}}{4}\,\hat{a}^\dagger+\ldots\right).
\end{equation}
\end{subequations}
Assuming the frequencies of different terms to be sufficiently different, we neglect the interference between them and treat each term of the form (\ref{eq:H1perturbation}) separately (we return to this point at the end of this section).

Using the standard assumption that the full system-bath density matrix remains factorizable at all times, $\hat\rho(t)\otimes\hat{\rho}_\text{qp}$ (that is, the incident quasiparticles have no memory of the qubit state), where the quasiparticle density matrix,
\begin{equation}\label{eq:rhoqp}
\hat{\rho}_\text{qp}=\prod_{\iota=L,R}\prod_k\left\{f(\epsilon_{\iota,k})\,\hat\gamma^\dagger_{\iota,k}\hat\gamma_{\iota,k}
+[1-f(\epsilon_{\iota,k})]\,\hat\gamma_{\iota,k}\hat\gamma_{\iota,k}^\dagger\right\},
\end{equation}
remains unchanged during the evolution, we pass to the interaction representation with respect to the Hamiltonian $\hat{H}_1$, writing $\hat\rho(t)=e^{-i\hat{H}_Kt}\,\hat{\tilde{\rho}}(t)\,e^{i\hat{H}_Kt}$. In the second order of the perturbation theory, the slow matrix $\hat{\tilde{\rho}}(t)$ satisfies the equation
\begin{equation}\label{eq:BornHamiltonian}
\frac{d\hat{\tilde{\rho}}(t)}{dt}=-\int_{-\infty}^t\Tr_\text{qp}\left\{
[\hat{\tilde{H}}_1(t),[\hat{\tilde{H}}_1(t'),\hat{\tilde{\rho}}(t')\otimes\hat\rho_\text{qp}]]\right\},
\end{equation}
where $\hat{\tilde{H}}_1(t)=e^{i\hat{H}_0t}\,\hat{H}_1(t)\,e^{-i\hat{H}_0t}$. Making the Markovian approximation, $\hat{\tilde{\rho}}(t')\approx\hat{\tilde{\rho}}(t)$, we take $\hat{\tilde{\rho}}(t)$ out of the integral, which now can be evaluated explicitly.

To determine the time dependence of $\hat{\tilde{H}}_1(t)$ explicitly, we pass to the basis $\{|\psi_l\rangle\}$ of eigenstates of $\hat{H}_K$, $\hat{H}_K|\psi_l\rangle=\varepsilon_l|\psi_l\rangle$ (for the sake of compactness, we momentarily suppress the parity index $\sigma$, so summations over~$l$ run over all eigenstates) and represent the operator $\hat{A}$ in terms of its matrix elements $A_{ll'}\equiv\langle\psi_{l}|\hat{A}|\psi_{l'}\rangle$:
\begin{equation}
e^{i\hat{H}_Kt}\hat{A}e^{-i\hat{H}_Kt}=\sum_{l,l'}
A_{ll'}e^{i(\varepsilon_{l}-\varepsilon_{l'})t}
|\psi_{l}\rangle\langle\psi_{l'}|.
\end{equation}
This expression, together with the oscillating factors $e^{\pm{i}\Omega{t}}$ in Eq.~\eqref{eq:H1perturbation}, determines the time dependence of the qubit part of $\hat{\tilde{H}}_1(t)$, while the time dependence of the quasiparticle operators is determined straightforwardly by~$\hat{H}_\text{qp}$.
The standard next step would be to substitute $\hat{\tilde{H}}_1(t)$ in Eq.~\eqref{eq:BornHamiltonian} and to make the secular approximation, which consists of neglecting all terms in Eq.~\eqref{eq:BornHamiltonian} that oscillate at non-zero frequencies. Here we make this approximation only partially, in the same spirit as in Ref.~\cite{Gasparinetti2013}: we neglect fast terms rotating as $e^{\pm2i\Omega{t}}$, but keep those proportional to $e^{i(\varepsilon_{l}-\varepsilon_{l'})t}$. We also note that the definition~(\ref{eq:Sqpdef}) implies
\begin{subequations}\begin{align}
&\Re\int_0^\infty\langle\hat{\cal I}(-t)\,\hat{\cal I}(0)\rangle\,e^{-i\omega{t}}\,dt
=\frac12 S_\text{qp}(\omega),\\
&\Re\int_0^\infty\langle\hat{\cal I}(0)\,\hat{\cal I}(-t)\rangle\,e^{-i\omega{t}}\,dt
=\frac12 S_\text{qp}(-\omega),
\end{align}\end{subequations}
and neglect the imaginary parts.
Indeed, terms originating from the imaginary parts (Lamb shifts) result in Hamiltonian perturbations of the form $\propto{i}[\hat{A}\hat{A}^\dagger,\hat\rho]$ in the master equation. The strongest term, $\propto{i}[\hat{a}^\dagger\hat{a},\hat\rho]$, can be corrected by choosing an appropriate drive frequency; higher-order terms, containing higher powers of $\hat{a}^\dagger\hat{a}$, are small by virtue of the assumption $\varphi_a|\alpha|\ll1$.

Finally, we pass back to the Schr\"odinger representation, and obtain the following equation for the matrix elements of $\hat\rho(t)$:
\begin{align}
\frac{d\rho_{ll'}}{dt}={}& -i(\varepsilon_l-\varepsilon_{l'})\rho_{ll'}
\nonumber\\ &{}+\sum_{l_1l_2}
\frac{S_\text{qp}(\Omega+\varepsilon_{l_2}-\varepsilon_{l'})}2\,A_{ll_1}\rho_{l_1l_2}A^\dagger_{l_2l'}
\nonumber\\ &{}+\sum_{l_1l_2}
\frac{S_\text{qp}(\Omega-\varepsilon_{l}+\varepsilon_{l_1})}2\,A_{ll_1}\rho_{l_1l_2}A^\dagger_{l_2l'}
\nonumber\\ &{}-\sum_{l_1l_2}
\frac{S_\text{qp}(\Omega-\varepsilon_{l_1}+\varepsilon_{l_2})}2\,A^\dagger_{ll_1}A_{l_1l_2}\rho_{l_2l'}
\nonumber\\ &{}-\sum_{l_1l_2}
\frac{S_\text{qp}(\Omega+\varepsilon_{l_1}-\varepsilon_{l_2})}2\,\rho_{ll_1}A^\dagger_{l_1l_2}A_{l_2l'}
\nonumber\\ &{}+\sum_{l_1l_2}
\frac{S_\text{qp}(-\Omega+\varepsilon_{l_2}-\varepsilon_{l'})}2\,A^\dagger_{ll_1}\rho_{l_1l_2}A_{l_2l'}
\nonumber\\ &{}+\sum_{l_1l_2}
\frac{S_\text{qp}(-\Omega-\varepsilon_{l}+\varepsilon_{l_1})}2\,A^\dagger_{ll_1}\rho_{l_1l_2}A_{l_2l'}
\nonumber\\ &{}-\sum_{l_1l_2}
\frac{S_\text{qp}(-\Omega-\varepsilon_{l_1}+\varepsilon_{l_2})}2\,A_{ll_1}A^\dagger_{l_1l_2}\rho_{l_2l'}
\nonumber\\ &{}-\sum_{l_1l_2}
\frac{S_\text{qp}(-\Omega+\varepsilon_{l_1}-\varepsilon_{l_2})}2\,\rho_{ll_1}A_{l_1l_2}A^\dagger_{l_2l'}.
\label{eq:nonsecular_master}
\end{align}
Since the frequency $\Omega > 0$ is a few GHz, while the energy scale of $\varepsilon_l$ is in the MHz range, in lines 2--5 of this equation one can approximate $S_\text{qp}(\Omega+\ldots)\approx{S}_\text{qp}(\Omega)$ (thus falling into the case studied in Ref.~\cite{Trushechkin2021}), and then these four lines wrap into the matrix element of ${S}_\text{qp}(\Omega)\,\mathcal{D}[\hat{A}]\hat\rho$, leading to Eqs.~\eqref{eq:kappa_minus} and \eqref{eq:kappa_plus} for $\hat{A}$ of Eqs.~\eqref{eq:A=a} and \eqref{eq:A=adag}, respectively. The last four lines contain ${S}_\text{qp}$ at negative frequencies, so they are suppressed by the Boltzmann factor. They can be wrapped into ${S}_\text{qp}(-\Omega)\,\mathcal{D}[\hat{A}^\dagger]\hat\rho$ if one neglects the Kerr qubit energies in the argument of $S_\text{qp}$; however, the typical frequency scale of $S_\text{qp}(\omega)$ at negative frequencies is given by the temperature~$T$, so such approximation requires a stronger condition $|\varepsilon_l-\varepsilon_{l'}|\ll{T}$.

Non-oscillating terms in the expansion of $\sin(\hat\varphi/2)$ produce a perturbation of the form $\hat{H}_1 = \hat{\cal I}\otimes\hat{A}$ with $\hat{A}=\hat{A}^\dagger=-\sin(\varphi_\text{bias}/2)(|\varphi_a|^2/4)\hat{a}^\dagger\hat{a}$ to the leading order in $|\varphi_a|\alpha$. The same perturbative treatment as above results in an equation similar to Eq.~(\ref{eq:nonsecular_master}), which can be obtained from Eq.~(\ref{eq:nonsecular_master}) by omitting the last four lines (to avoid double counting) and setting $\Omega\to0$. If we make the replacement~(\ref{eq:logprecision}) in all terms, the sums wrap into ${S}_\text{qp}(K\alpha^2)\mathcal{D}[\hat{A}]\hat\rho$ leading to Eq.~(\ref{eq:kappa_phi}).

Let us now estimate different contributions for the sample described in Ref.~\cite{Grimm2020}. The qubit is hosted in an aluminum superconducting loop (with gap $\Delta= \SI{200}{\micro\electronvolt}$) containing a small Josephson junction with the Josephson energy $E_J/(2\pi)=22\:\mbox{GHz}$ and three larger junctions with the Josephson energy $E_J'/(2\pi)=200\:\mbox{GHz}$, pierced by a magnetic flux $\Phi=0.26$ (in the units of the superconducting flux quantum) and maintained at temperature $T=18\:\mbox{mK}$. 
The oscillator frequency $\omega_a=2\pi\times6\:\mbox{GHz}$ and the drive frequency $\omega_\text{d}\approx2\omega_a$. 
The phase difference across the small junction is described by Eq.~(\ref{eq:phij}) with $\varphi_\text{bias}=-1.31$, $\varphi_a=0.20$, $\varphi_\text{d}=0.06$,  while the phase difference on each of the large junctions is parametrized by $\varphi_\text{bias}'=0.11$, $\varphi_a'=0.07$, $\varphi_\text{d}'=0.02$
($\varphi_a$ and $\varphi_a'$ can be found from the oscillator frequency and the Josephson potential). 
The parameters of the Kerr qubit are $K=2\pi\times6.7\:\mbox{MHz}$, $\alpha^2=2.5$.

Taking the perturbation~(\ref{eq:A=a}), we obtain $\kappa_-/(2\pi) = x_\text{qp}\times8.1\:\mbox{GHz}$, the main contribution coming from quasiparticle tunneling across the three large junctions and  only 17\% being due to tunneling across the small junction. The contribution to $\kappa_+$ from the Hermitian conjugate of the perturbation~(\ref{eq:A=a}) is smaller by a factor $S_\text{qp}(-\omega_a)/S_\text{qp}(\omega_a)$, which for the thermal distribution of quasiparticles amounts to $e^{-\omega_a/T}\sim10^{-7}$ for all junctions. The contribution to $\kappa_+$ from the perturbation~(\ref{eq:A=adag}) is more significant due to the absence of a thermal factor: for each junction~$j$, it is smaller than $\kappa_-$ by the factor $(\varphi_{\text{d},j}/2)^2\tan^2(\varphi_{\text{bias},j})$, which gives $\kappa_+\sim10^{-4}\kappa_-$ with the main contribution coming from the small junction. The dephasing rate $\kappa_\phi/(2\pi) = x_\text{qp}\times50\:\mbox{MHz}$ is also dominated by the small junction.

In the experiment~\cite{Grimm2020}, the measured qubit error rates were reproduced by the solution of Eq.~(\ref{eq:masterequation}) with $\kappa_-/(2\pi)\sim10\:\mbox{kHz}$ and rather large $\kappa_+/\kappa_-\sim0.04$, $\kappa_\phi/\kappa_-\sim0.02$. 
If we assume that the photon loss is entirely due to quasiparticles, it would require the concentration $x_\text{qp}\sim10^{-6}$, which is rather high, but still realistic.
However, the quasiparticles would not be able to reproduce the observed values of $\kappa_+,\kappa_\phi$. Other dissipation mechanisms are needed to explain these values.

A subtle detail about the experimental realization of the Kerr qubit~\cite{Grimm2020} is that the detuning $\omega_\text{d}-2\omega_a=-4.4\:\mbox{MHz}\times2\pi$ is not large enough compared to the typical spacing between the energy levels of the Kerr Hamiltonian, $\sim{K}\alpha^2$. Then, the perturbations proportional to $\hat{a}$ and $\hat{a}^\dagger$ rotate at close frequencies, and one cannot treat them as two separate perturbations of the form (\ref{eq:H1perturbation}), as we briefly mentioned in Sec. \ref{ssec:quasiparticle-induced}. Here we prefer not to study the case of a general detuning; we just compare the limit of large detuning, which results in two separate dissipators $\kappa_-\mathcal{D}[\hat{a}]+\kappa_+\mathcal{D}[\hat{a}^\dagger]$, and the limit of zero detuning which again has the form (\ref{eq:H1perturbation}) and yields a single coherent dissipator $\mathcal{D}\left[\sqrt{\kappa_-}\,\hat{a}+\sqrt{\kappa_+}\,\hat{a}^\dagger\right]$. We check numerically in Appendix~\ref{app:numerics} that these two extremes lead to very similar results. Thus, from the practical point of view, one can still use the phenomenological master equation~(\ref{eq:masterequation}).

\section{Dissipators for the dissipative qubit}
\label{sec:DerivationDissip}

In this section, we address the (rather unconventional) task of deriving the error dissipators perturbatively in the system-bath coupling when the uncoupled system's dynamics is not Hamiltonian, but dissipative.
In the conventional case when the unperturbed qubit dynamics is Hamiltonian (as in the previous section), tracing out the quasiparticles in the Born-Markov approximation yields the dissipators in terms of the quasiparticle correlator $S_\text{qp}(\omega)$ at real frequencies, determined by transitions between energy levels of the unperturbed qubit; these finite transition frequencies regularize most of the logarithmically divergent terms in the dephasing dissipator.
The unperturbed dissipative qubit already has no energy levels, but still has non-trivial dynamics; hence, it is not clear \emph{a priori}, at what frequencies the quasiparticle correlator should enter the error dissipators.
Thus, we are obliged to revisit the whole derivation scheme for master equation, starting from a Lindbladian unperturbed qubit dynamics and Hamiltonian qubit-quasiparticle coupling.

We start from the Liouville-von Neumann equation for the density matrix $\rho_\text{tot}(t)$ of the total system ``qubit oscillator + quasiparticles,''
\begin{equation}
\frac{d\hat\rho_\text{tot}(t)}{dt} = \mathcal{L}_0\hat\rho_\text{tot}(t)
-i[\hat{H}_\text{qp},\hat\rho_\text{tot}(t)]-i[\hat{H}_1(t),\hat\rho_\text{tot}(t)],
\end{equation}
where $\mathcal{L}_0=\kappa_2\mathcal{D}[\hat{a}^2-\alpha^2]$ is the familiar Lindbladian of the dissipative qubit, while $\hat{H}_\text{qp}$ and $\hat{H}_1(t) = \hat{\cal I}\otimes\hat{A}(t)$ are given by Eqs.~(\ref{eq:H0Kqp}) and (\ref{eq:H1perturbation}), respectively.
The passage to the interaction representation with respect to the perturbation $\hat{H}_1(t)$ now has the form
\begin{equation}
\hat\rho_\text{tot}(t)=e^{\mathcal{L}_0t}\left(e^{-i\hat{H}_\text{qp}t}\hat{\tilde\rho}_\text{tot}(t)e^{i\hat{H}_\text{qp}t}\right),
\end{equation}
and leads to the following equation of motion for $\hat{\tilde\rho}_\text{tot}(t)$:
\begin{align}
\frac{d\hat{\tilde\rho}_\text{tot}(t)}{dt} ={}&{} {-i}e^{-\mathcal{L}_0t}
\left[e^{i\hat{H}_\text{qp}t}\hat{H}_1(t)e^{-i\hat{H}_\text{qp}t}
\,,\,e^{\mathcal{L}_0t}\hat{\tilde\rho}_\text{tot}(t)\right]\nonumber\\
\equiv{}&{} \tilde{\mathcal{L}}_1(t)\hat{\tilde\rho}_\text{tot}(t).
\end{align}
As usual, we assume the full system-bath density matrix to remain factorizable at all times, $\hat{\tilde\rho}_\text{tot}(t)=\hat{\tilde\rho}(t)\otimes\hat{\rho}_\text{qp}$, with the quasiparticle density matrix $\hat{\rho}_\text{qp}$ given by Eq.~(\ref{eq:rhoqp}).
Then, in the second order of the perturbation theory, the slow matrix $\hat{\tilde{\rho}}(t)$ satisfies the equation
\begin{equation}\label{eq:BornLindbladian}
\frac{d\hat{\tilde{\rho}}(t)}{dt}=\int_{-\infty}^tdt'\,\Tr_\text{qp}\left\{
\tilde{\mathcal{L}}_1(t)\tilde{\mathcal{L}}_1(t')\left[\hat{\tilde{\rho}}(t')\otimes\hat\rho_\text{qp}\right]\right\},
\end{equation}
which so far looks quite analogous to Eq.~(\ref{eq:BornHamiltonian}) for the Hamiltonian case.

To write the time dependence explicitly, we assume that the Lindbladian $\mathcal{L}_0=\kappa_2\mathcal{D}[\hat{a}^2-\alpha^2]$ in zeroth-order approximation has a complete biorthogonal set of left and right eigenvectors.%
\footnote{
We are not aware of a rigorous mathematical proof of the existence of a complete set of eigenvectors (which is not guaranteed \textit{a priori} for a non-Hermitian operator $\mathcal{L}_0$). Still, numerical diagonalization in a truncated basis does not show any sign of the opposite.
}
Since the superoperator $\mathcal{L}_0$ commutes with the left and right parity, $(-1)^{\hat{a}^\dagger\hat{a}}\hat\rho=\sigma\hat\rho$ and $\hat\rho(-1)^{\hat{a}^\dagger\hat{a}}=\sigma'\hat\rho$, respectively, its eigenvalues $\lambda_m^{\sigma\sigma'}$ can be labeled by the two parities $\sigma,\sigma'=\pm$ and an integer $m\geq{0}$, such that $\lambda_0^{\sigma\sigma'}=0$. The left and right eigenvectors, $\hat\varsigma_m^{\sigma\sigma'}$ and $\hat\varrho_m^{\sigma\sigma'}$, are assumed to form a complete biorthogonal set:%
\footnote{
We define the left eigenvectors $\hat\varsigma_m^{\sigma\sigma'}$ in such a way that they enter Eqs.~(\ref{eq:Lindbladian_eigenvectors}) without Hermitian conjugation to compactify the notations. Then the superscripts $\sigma\sigma'$ at $\hat\varsigma_m^{\sigma\sigma'}$ should be understood as a label indicating the corresponding eigenvalue, while the left (right) parity operation acts as
$(-1)^{\hat{a}^\dagger\hat{a}}\hat\varsigma_m^{\sigma\sigma'}=\sigma'\hat\varsigma_m^{\sigma\sigma'}$ and $\hat\varsigma_m^{\sigma\sigma'}(-1)^{\hat{a}^\dagger\hat{a}}=\sigma\hat\varsigma_m^{\sigma\sigma'}$.
}
\begin{subequations}\label{eq:Lindbladian_eigenvectors}\begin{align}
&\mathcal{L}_0\hat\varrho_m^{\sigma\sigma'}=\lambda_m^{\sigma\sigma'}\hat\varrho_m^{\sigma\sigma'},\\
&\Tr\left\{\hat\varsigma_m^{\sigma\sigma'}\mathcal{L}_0\hat\rho\right\}=
\lambda_m^{\sigma\sigma'}\Tr\left\{\hat\varsigma_m^{\sigma\sigma'}\hat\rho\right\}\quad\forall\hat\rho,\\
&\Tr\{\hat\varsigma_m^{\sigma\sigma'}\hat\varrho_{m'}^{\sigma\sigma'}\}=\delta_{mm'},\\
&\sum_{m,\sigma,\sigma'}\hat\varrho_m^{\sigma\sigma'}\Tr\{\hat\varsigma_m^{\sigma\sigma'}\hat\rho\}
=\hat\rho\quad\forall\hat\rho.
\label{eq:completeness}
\end{align}\end{subequations}
For $m=0$, the right eigenvectors are simply $\hat\varrho_0^{\sigma\sigma'}=|\mathcal{C}_\alpha^\sigma\rangle\langle\mathcal{C}_\alpha^{\sigma'}|$, while the four left eigenvectors $\hat\varsigma_0^{\sigma\sigma'}$ (also called the invariants of the dynamics, since $(\partial/\partial{t})\Tr\{\hat\varsigma_0^{\sigma\sigma'}\hat\rho\}=0$) can be found in Ref.~\cite{Guillaud2023} and in Appendix~\ref{app:perturbations}.

We define the matrix elements of the superoperators, corresponding to multiplication by an arbitrary operator~$\hat{O}$ from the left and right (for the sake of compactness, we momentarily suppress the parity indices $\sigma,\sigma'$, so summations over~$m$ run over all eigenvectors),
\begin{subequations}\begin{align}
&\hat{O}\hat\varrho_m = \sum_{m'}\overrightarrow{O}_{m'm}\hat\varrho_{m'},\quad
\overrightarrow{O}_{m'm}=\Tr\left\{ \hat\varsigma_{m'}\hat{O}\hat\varrho_m\right\},\\
&\hat\varrho_m\hat{O} = \sum_{m'}\overleftarrow{O}_{m'm}\hat\varrho_{m'},\quad
\overleftarrow{O}_{m'm}=\Tr\left\{ \hat\varrho_m\hat{O}\hat\varsigma_{m'}\right\},
\end{align}\end{subequations}
and expand the density matrix in this basis:
\begin{equation}
\hat{\tilde{\rho}}(t)=\sum_m\tilde{r}_m(t)\,\hat\varrho_m.
\end{equation}
We also introduce the quasiparticle structure factor in the time representation, $\tilde{S}_\text{qp}(t-t')\equiv\Tr\{\hat{\cal I}(t)\,\hat{\cal I}(t')\,\hat\rho_\text{qp}\}$. For the thermal $\hat\rho_\text{qp}$, given by Eq.~(\ref{eq:rhoqp}) with the distribution~(\ref{eq:fthermal}), it evaluates to
\begin{equation}
\tilde{S}_\text{qp}(t)=\frac{8E_J}{\pi}\sqrt{\frac{\Delta}{2\pi{T}}}\,
\frac{x_\text{qp}}{\sqrt{(t+i/T)(t-i0^+)}}.
\end{equation}
Then we can transform Eq.~(\ref{eq:BornLindbladian}) into an equation for the coefficients~$\tilde{r}_m(t)$:
\begin{widetext}
\begin{align}
\frac{d\tilde{r}_m(t)}{dt} ={}&{}-\sum_{m',m''}e^{(\lambda_{m''}-\lambda_m)t}\int_{-\infty}^tdt'\,
e^{(\lambda_{m'}-\lambda_{m''})(t-t')}\nonumber\\
{}&{}\qquad\times\left[\overrightarrow{A}_{mm'}(t)-\overleftarrow{A}_{mm'}(t)\right]
\left[\tilde{S}_\text{qp}(t-t')\,\overrightarrow{A}_{m'm''}(t')
- \tilde{S}_\text{qp}(t'-t)\,\overleftarrow{A}_{m'm''}(t')\right]
\tilde{r}_{m''}(t').
\label{eq:expdiverge}
\end{align}
\end{widetext}
Here one can observe the crucial difference from the Hamiltonian case: since the eigenvalues $\lambda_m$ have non-zero real parts, the time integral may diverge exponentially. In the Hamiltonian case, the exponential factor $e^{(\lambda_{m'}-\lambda_{m''})(t-t')}$ is purely oscillatory; combined with the oscillating or constant $\overrightarrow{A}_{m'm''}(t')$, $\overleftarrow{A}_{m'm''}(t')$, and with the decaying $\tilde{S}_\text{qp}(t-t')\sim1/|t-t'|$, this is sufficient to ensure the convergence of the integral and to justify the Markovian approximation, which essentially means that the integral is dominated by $\tilde{r}_{m''}(t'\approx{t})$. Exponential divergence in the dissipative case means that the values $\tilde{r}_{m''}(t')$ in the remote past are more important than those at $t'\approx{t}$, so a Markovian master equation cannot be derived. In simple terms, it makes no sense to study weak dissipative perturbations of an already strongly dissipative dynamics. This problem can be bypassed in two cases.   

First, if the relaxation rates $|\Re\lambda_m|$ are small compared to the oscillation frequencies [e.~g., to $\Omega$ in Eq.~(\ref{eq:H1perturbation})], one can cut off the time integral in Eq.~(\ref{eq:expdiverge}) at times ${t}-t'\sim\tau_*$ such that $1/\Omega\ll\tau_*\ll1/|\Re\lambda_{m''}|$, and make the Markovian approximation. The resulting master equation describes the dynamics, coarse-grained in time on the scale~$\tau_*$, so the dissipation contained in $\mathcal{L}_0$ and the dissipation due to the perturbation $\hat{H}_1$ are treated on equal footing. This is equivalent to simply neglecting the dissipative part of $\mathcal{L}_0$ and effectively deriving the dissipators due to $\hat{H}_1$ as in the Hamiltonian case. For the dissipative qubit, this gives the dissipators $\mathcal{D}[\hat{a}]$ and $\mathcal{D}[\hat{a}^\dagger]$ with the rates $\kappa_-$ and $\kappa_+$ given by Eqs.~(\ref{eq:kappa_minus}) and~(\ref{eq:kappa_plus}), the same as for the Kerr qubit.

Second, in the absence of fast oscillations, Eq.~(\ref{eq:expdiverge}) can be cast into a Markovian form for those $m''$ which have $\lambda_{m''}=0$, that is, those in the computational subspace of the dissipative qubit. Then the time integral converges exponentially for $\Re\lambda_{m'}<0$, and it is easy to see that it results in the quasiparticle current spectral density, taken at an imaginary frequency. Indeed, for real frequencies~$\omega$ we have (including the factor of~2 from spin)
\begin{align}
S_\text{qp}(\omega)\equiv{}&{}
\int_{-\infty}^\infty\langle\hat{\cal I}(t)\,\hat{\cal I}(0)\rangle\,{e}^{i\omega{t}}\,{dt}\nonumber\\
={}&{}\frac{8E_J}{\pi\Delta}
\int\limits_{-\infty}^\infty{}\!\!d\xi_k\,d\xi_{k'}\,
f(\epsilon_k)[1\!-\!f(\epsilon_{k'})]
\delta(\epsilon_k-\epsilon_{k^\prime}+\omega)\nonumber\\
={}&{}\frac{16E_J}\pi\int_0^\infty\frac{d\epsilon}{\sqrt\epsilon}\,
\frac{\theta(\epsilon+\omega)}{\sqrt{\epsilon+\omega}}\,f(\epsilon),
\label{eq:Sqpgeneral}
\end{align}
where we assume the quasiparticles to be dilute, so the occupations $f(\epsilon_k)\ll1$ and we replace $1-f(\epsilon_{k^\prime})\to1$, arriving at Eq. \eqref{eq:Sthermal}. For a decaying exponential,
\begin{align}
&\int_0^\infty\langle\hat{\cal I}(\pm{t})\,\hat{\cal I}(0)\rangle\,e^{-\Gamma{t}}\,dt\nonumber\\
{}&{}=\frac{4E_J}{\pi^2\Delta}\int_{-\infty}^\infty{}d\xi_k\,d\xi_{k'}\,
\frac{f(\epsilon_k)[1-f(\epsilon_{k'})]}%
{\mp{i}(\epsilon_k-\epsilon_{k'})+\Gamma}\nonumber\\
{}&{}=\frac{8E_J}\pi\int_0^\infty
\frac{d\epsilon\,f(\epsilon)}{\sqrt{\epsilon(\epsilon\pm{i}\Gamma)}}
=\frac{1}{2}\,S_\text{qp}(\pm{i}\Gamma),
\end{align}
where $S_\text{qp}(\omega)$ at complex $\omega$ is understood as the analytical continuation from the real positive semiaxis. For $\lambda_m=0$, the corresponding components of Eq.~(\ref{eq:expdiverge}) give the projected dynamics in the computational subspace. Restoring the left (right) parity indices $\sigma,\sigma'$, we find that the projected dephasing dissipator $\mathcal{P}_\|\kappa_\phi\mathcal{D}[\hat{a}^\dagger\hat{a}]\hat\rho_\|$ in Eq.~(\ref{eq:masterprojected}) is effectively replaced by a superoperator $\mathcal{L}_{\phi}\hat\rho_\|$, defined on the computational subspace~$S_\|$ by its action on the basis matrices $\hat\varrho_0^{\sigma\sigma'}$:
\begin{widetext}
\begin{align}
\mathcal{L}_{\phi}\hat\varrho_0^{\sigma\sigma'}={}&{}
\delta_{\sigma,-\sigma'}\hat\varrho_0^{\sigma\sigma'}\,\frac{|\varphi_a|^4}{32}\sin^2\frac{\varphi_\text{bias}}{2}\nonumber\\
&{}\times\sum_{m=0}^\infty
\Tr\left\{[\hat\varsigma_m^{\sigma\sigma'},\hat{a}^\dagger\hat{a}]\varrho_0^{\sigma\sigma'}\right\}
\left[S_\text{qp}(i\lambda_m^{\sigma\sigma'})\Tr\left\{\hat\varsigma_{0}^{\sigma\sigma'}\hat\varrho_m^{\sigma\sigma'}\hat{a}^\dagger\hat{a}\right\}
-S_\text{qp}(-i\lambda_m^{\sigma\sigma'})\Tr\left\{\hat\varsigma_{0}^{\sigma\sigma'}\hat{a}^\dagger\hat{a}\varrho_m^{\sigma\sigma'}\right\}\right].
\label{eq:Lphiprojected=}
\end{align}
\end{widetext}
Similarly to the Kerr qubit case, we notice that $-\lambda_m^{\sigma\sigma'}\sim\kappa_2\alpha^2\ll{T}$, and if we take all $S_\text{qp}(\pm{i}\lambda_m^{\sigma\sigma'})\approx S_\text{qp}(\kappa_2\alpha^2)$ with logarithmic precision, then the completeness relation~(\ref{eq:completeness}) results in $\mathcal{L}_{\phi}=\mathcal{P}_\|\kappa_\phi\mathcal{D}[\hat{a}^\dagger\hat{a}]$ with $\kappa_\phi$ given by Eq.~(\ref{eq:kappa_phi}). Again, this argument does not work for the $m=0$ term with $\lambda_0^{\sigma\sigma'}=0$, so the divergence of $S_\text{qp}(\omega\to0)$ persists, and has to be handled as in Refs.~\cite{Catelani2012,Zanker2015}. However, the relative contribution of the $m=0$ term is exponentially small; evaluating the contributions from $m>0$ and $m=0$ separately, we find a rate proportional to $S_\text{qp}(\kappa_2\alpha^2)+(15/4)e^{-2\alpha^2}S_\text{qp}(0)$ at $\alpha\gg1$ (omitting all common factors).

Returning to Eq.~(\ref{eq:expdiverge}),
let us now focus on terms with $\lambda_m\neq0$. In fact, the growing exponential $e^{-\lambda_mt}$ disappears when one goes back from the interaction representation to the Schr\"odinger one. Then, the corresponding superoperator describes the leakage from the computational subspace, represented by $\mathcal{P}_\perp\mathcal{L}_1\hat\rho_\|$ in Eq.~(\ref{eq:rhoperp=}), which is thus well defined.

Let us now estimate different contributions for the sample described in Ref.~\cite{Lescanne2020}. It includes two small Josephson junctions with $E_J/(2\pi)=90\:\mbox{GHz}$ and an inductor consisting of five  larger junctions with $E_J'/(2\pi)=225\:\mbox{GHz}$, all made of aluminum with $\Delta=\SI{200}{\micro \electronvolt}$ and maintained at temperature $T=10\:\mbox{mK}$. These three elements (the two small junctions and the inductor) are all connected in parallel and have the same phase drop~$\varphi$, except one of the small junctions which is also subject to a static flux bias of a half flux quantum, thus having the phase drop $\varphi+\pi$. 
The qubit oscillator frequency $\omega_a=2\pi\times8.0\:\mbox{GHz}$ and the drive frequency $\omega_\text{d}=2\pi\times11.2\:\mbox{GHz}$.
The main dissipator defining the cat qubit has the rate $\kappa_2=2\pi\times40\:\mbox{kHz}$, and $\alpha^2$ was varied in the range $0-7$.
The phase difference across the small junctions is described by Eq.~(\ref{eq:phij}) with $\varphi_\text{bias}=0$ or~$\pi$, $\varphi_a\approx\varphi_\text{d}\approx0.03$ (extracted from the measured values of $\kappa_2$ and frequency shifts of the qubit oscillator and of the buffer oscillator providing the strong dissipation), while each of the large junctions has $\varphi_\text{bias}'=0$ and $\varphi_a'\approx\varphi_\text{d}'=\varphi_\text{d}/5$.
Taking the perturbation~(\ref{eq:A=a}), we obtain $\kappa_-/(2\pi) = x_\text{qp}\times0.27\:\mbox{GHz}$, the main contribution ($\sim70\%$) coming from quasiparticle tunneling across the unbiased small junction.
$\kappa_+\sim(2\times10^{-4})\kappa_-$ and $\kappa_\phi\sim(3\times10^{-3})\kappa_-$ for $\alpha^2=7$ are determined by the $\pi$-biased small junction. Again, $\kappa_+$ comes from the drive perturbation~(\ref{eq:A=adag}) rather than thermal quasiparticle population. 

In Ref.~\cite{Lescanne2020}, the rate $\kappa_-=2\pi\times53\:\mbox{kHz}$ was measured (no reliable values for $\kappa_+$ and $\kappa_\phi$ could be extracted). To produce such a high rate, one needs the quasiparticle concentration $x_\text{qp}\sim10^{-4}$. This value seems to be unrealistically high, so that the photon loss is likely to be dominated by other mechanisms. (It should be noted, though, that while the Josephson junctions were made of aluminium, much of the circuit was made of niobium. The latter has a larger gap, so the aluminium islands would attract all quasiparticles generated in the niobium part of the circuit. This could possibly lead to an unusually high effective $x_\text{qp}$.)

\section{Quasiparticle kinetics}
\label{sec:Kinetics}

\subsection{Qualitative discussion}\label{ssec:qualitativekinetics}

So far, we have taken for granted that the quasiparticle energy distribution is thermal, Eq.~(\ref{eq:fthermal}), with the temperature~$T$ fixed by the phonon bath. However, the system is subject to a microwave drive, so the quasiparticles absorb energy from the driving field, and their energy distribution can deviate from the thermal one. In fact, the quasiparticle distribution in the stationary state is determined by the competition between energy absorption from the drive and thermalization with phonons. 
In this section, we investigate this issue using the approach of Refs.~\cite{Catelani2019, Fisher2023, Fisher2024}.

Quasiparticles interact with the drive photons when they tunnel through Josephson junctions, while they interact with phonons anywhere inside each island. This means that the interaction with phonons, a bulk effect, should dominate over the interaction with photons, a surface effect, in large enough islands. In small islands, if the phonon emission is not efficient enough, the hot quasiparticles can escape to neighboring large islands.
Thus, we expect that photon absorption by thermal quasiparticles due to the perturbation~(\ref{eq:perturbationsin}), expanded to the linear order in the oscillating terms $\propto{e}^{-i\omega_\text{d}t},e^{-i\omega_at}$ in Eq.~(\ref{eq:phij}), creates weak replicas of the main thermal population at energies $\epsilon$ such that $0<\epsilon-\omega_\text{d}\sim{T}$ and $0<\epsilon-\omega_a\sim{T}$. (We remind that we measure $\epsilon$ from the superconducting gap $\Delta$.) In the following, we estimate the strength of these replicas, neglecting the depletion of the main thermal population at $\epsilon\sim{T}$.

Among the error rates that we have studied in the previous sections, it is the photon absorption rate $\kappa_+$ that is most sensitive to excess quasiparticle population at high energies. Indeed, we have seen that the leading-order contribution is $\kappa_+\propto{S}_\text{qp}(-\omega_a)\propto{f}(\omega_a)$. This could make one think that the replica at $\epsilon\approx\omega_a$ may give an important contribution to ${S}_\text{qp}(-\omega_a)$. However, appearance of a quasiparticle with energy $\epsilon\approx\omega_a$ is necessarily a consequence of a photon loss error in the qubit. Thus, this contribution cannot be associated with the dissipator $\kappa_+\mathcal{D}[\hat{a}^\dagger]$, independent from $\kappa_-\mathcal{D}[\hat{a}]$. Therefore, we will estimate the contribution to $\kappa_+$ from the replica at $\epsilon\approx\omega_\text{d}>\omega_a$ only.

\subsection{Inelastic and elastic rates}
Here we calculate the rates of different quasiparticle transitions, which are needed to determine the stationary quasiparticle distribution. 

Fermi's golden rule with the perturbation \eqref{eq:HJqp} and the phase expansion \eqref{eq:phij} yields the probability per unit time to fill a given state $k$ with a given spin on an island~$\iota$ by any quasiparticle absorbing a photon in the form of a sum over all junctions~$j$ involving the island~$\iota$:
\begin{align}
\Gamma_\text{abs}(\epsilon_k) = {}&{}\sum_{j\in{\iota}}\sum_{k'}f(\epsilon_{k'})\,2\pi\mathcal{T}_{j,kk'}^2
\cos^2\frac{\varphi_{\text{bias},j}}{2}{}\nonumber\\
{}&{}\times\left[\frac{|\varphi_{\text{d},j}|^2}{4}\,\delta(\epsilon_{k'}+\omega_\text{d}-\epsilon_k)
\right.\nonumber\\
{}&{}\quad +\left. \langle\hat{a}^\dagger\hat{a}\rangle\,\frac{|\varphi_{a,j}|^2}{4}\,\delta(\epsilon_{k'}+\omega_\text{a}-\epsilon_k)
\right]\nonumber\\
= {}&{} \sum_{j\in{\iota}}\frac{E_{Jj}\delta_\iota}\pi\,\cos^2\frac{\varphi_{\text{bias},j}}{2}{}\nonumber\\
{}&{}\times \left[\frac{|\varphi_{\text{d},j}|^2f(\epsilon_k-\omega_\text{d})}{\sqrt{2\Delta(\epsilon_k-\omega_\text{d})}}
+ \frac{|\varphi_{a,j}|^2\langle\hat{a}^\dagger\hat{a}\rangle f(\epsilon_k-\omega_a)}{\sqrt{2\Delta(\epsilon_k-\omega_a)}}\right].
\label{eq:absRate}
\end{align}
%
where each term in the square bracket is present only for such $\epsilon_k$ that the argument of the square root is positive. The average $\langle\hat{a}^\dagger\hat{a}\rangle$ is over the stationary density matrix of the qubit, $\langle\hat{a}^\dagger\hat{a}\rangle\approx \alpha^2$ up to exponentially small terms. The mean-level spacing $\delta_\iota=1/(\nu_0V_\iota)$ is defined by Eq.~(\ref{eq:meanlevelspacing}), and is inversely proportional to the island's volume~$V_\iota$. At the same time, $E_{Jj}$ is proportional to the area of the $j$th junction. This matches the discussion of the previous paragraph: a quasiparticle living in a large island can absorb a photon only when it comes close to the junction, hence the surface-to-volume ratio.

To describe the quasiparticle thermalization with phonons, we adopt the standard model of electrons coupled to acoustic phonons~\cite{Kaplan1976,Chang1977} in which the effective electron-phonon coupling is $\alpha^2\,F(\omega)\propto\omega^2$ for the phonon frequency~$\omega$.
Various material parameters entering the phonon emission rate can be conveniently wrapped into a single coefficient. To relate to experimentally measured values, available in the literature, one can use the time $\tau_0$, such that the phonon emission rate (i.~e., the golden-rule probability per unit time for a quasiparticle initially in a given state to leave this state by emitting a phonon) by a quasiparticle of energy $\epsilon_k\gg\Delta$ is given by~\cite{Kaplan1976}
\begin{equation}
\Gamma_\text{em}(\epsilon_k\gg\Delta)=\frac{1}{3\tau_0}\,\frac{\epsilon_k^3}{T_c^3},
\end{equation}
where $T_c$ is the superconductor's critical temperature.
Alternatively, one may use the coefficient $\Sigma$, which controls the energy exchange between electrons and phonons for the material in the normal state: the power per unit volume transferred from electrons to phonons, kept at temperatures $T_\text{e}$ and $T_\text{ph}$, respectively, is given by $\Sigma(T_\text{e}^5 - T_\text{ph}^5)$~\cite{Roukes1985, Kautz1993, Wellstood1994}. The two coefficients are related:
\begin{equation}\label{eq:Sigmatau0}
\frac{1}{\tau_0}=\frac{\Sigma{T}_c^3}{48\,\zeta(5)\,\nu_0}.
\end{equation}
In this model, the phonon emission rate for a quasiparticle with energy~$\epsilon_k\ll\Delta$ is given by~\cite{Kaplan1976,Catelani2019}
\begin{equation}\label{eq:emRate}
\Gamma_\text{em}(\epsilon_k\ll\Delta)=\frac{8}{315\,\zeta(5)}\,\frac{\Sigma\epsilon_k^{7/2}}{\sqrt{2\Delta}\,\nu_0}
=\frac{128}{105}\,\frac{\epsilon_k^{7/2}}{\sqrt{2\Delta}\,T_c^3}\,\frac{1}{\tau_0}.
\end{equation}
The values of $\Sigma$ and $\tau_0$ for aluminum, found in the literature, are not very consistent among themselves: the available values of $\Sigma = (0.2-0.3)\times10^9\:\text{W}/(\text{m}^2\cdot\text{K}^5)$~\cite{Kautz1993, Meschke2004} yield $\tau_0\sim \SI{1}{\micro \second}$ according to Eq.~(\ref{eq:Sigmatau0}), while Refs.~\cite{Chi1979,Moody1981} report $\tau_0\sim100\:\mbox{ns}$. For the estimates below, we use the most recent information, namely, $\Sigma = 0.3\times10^9\:\text{W}/(\text{m}^2\cdot\text{K}^5)$ \cite{Meschke2004}.

As the smallest islands in the dissipative Kerr qubit of Ref.~\cite{Lescanne2020} are connected by junctions with $\varphi_{\text{bias},j}=0$, we use the Hamiltonian~(\ref{eq:HJqp}) with the next-to-leading order expansion of the Bogolyubov coefficients to find the rate of elastic tunneling to neighboring islands (since the leading order vanishes). The golden rule then gives the probability per unit time for a given quasiparticle on an island~$\iota$ to escape this island via any junction:
\begin{align}
\Gamma_\text{esc} = {}&{}\sum_{j\in{\iota}}\sum_{k'}2\pi\mathcal{T}_{j,kk'}^2
\,\delta(\epsilon_{k'}-\epsilon_k)\nonumber\\
{}&{}\times\left(\sin^2\frac{\varphi_{\text{bias},j}}{2}
+\frac{\xi_k^2}{\xi_k^2+\Delta^2}\,\cos^2\frac{\varphi_{\text{bias},j}}{2}\right){}\nonumber\\
\approx {}&{} \sum_{j\in{\iota}}\frac{4E_{Jj}\delta_\iota}{\pi\sqrt{2\Delta\epsilon_k}}\left(\sin^2\frac{\varphi_{\text{bias},j}}{2}
+\frac{2\epsilon_k}\Delta\,\cos^2\frac{\varphi_{\text{bias},j}}{2}\right).
\label{eq:escRate}
\end{align}

\subsection{Hot quasiparticle population}
As discussed in Sec. \ref{ssec:qualitativekinetics}, the quasiparticles are more likely to overheat in small islands.
In both experimental realizations~\cite{Grimm2020,Lescanne2020}, the smallest islands are located inside the chain of $N$ large junctions ($N=3$ in Ref.~\cite{Grimm2020} and $N=5$ in Ref.~\cite{Lescanne2020}). Thus, we focus on quasiparticle overheating on islands $\iota=1,\ldots,N-1$, assuming them to be identical. The two islands $\iota=0,N$ terminating the chain are assumed to be large, so the quasiparticle overheating on these islands is neglected. All $N$ junctions are also assumed to be identical, with the Josephson energy $E_J'$. Denoting by $\tilde{f}_\iota(\epsilon)\equiv{f}(\epsilon)-f_T(\epsilon)$ the correction to the thermal distribution at high energies $\epsilon\approx\omega_\text{d},\omega_a$ due to photon absorption, we can write the kinetic equation as
\begin{align}
0 = {}&{}\Gamma_\text{abs}(\epsilon) - \Gamma_\text{em}(\epsilon)\,\tilde{f}_\iota(\epsilon)\nonumber\\
{}&{} + \Gamma_\text{esc}(\epsilon)
\left[\frac{\tilde{f}_{\iota+1}(\epsilon)+\tilde{f}_{\iota-1}(\epsilon)}{2}-\tilde{f}_\iota(\epsilon)\right]
\end{align}
with the boundary conditions $\tilde{f}_{\iota=0}=\tilde{f}_{\iota=N}=0$. We assume that only a small fraction of quasiparticles is excited, so we neglect the depletion of the main thermal population and evaluate  $\Gamma_\text{abs}(\epsilon)$ using Eq.~(\ref{eq:absRate}) with the thermal $f(\epsilon)=f_T(\epsilon)$, Eq.~(\ref{eq:fthermal}). Then the kinetic equation is straightforwardly solved  by expanding in the eigenfunctions of the discrete Laplacian with zero boundary conditions, $\sqrt{2/N}\sin(\pi{m}\,\iota/N)$, labeled by $m=1,\ldots,N-1$:
\begin{subequations}\begin{align}
&\tilde{f}_\iota(\epsilon) = \sum_{m=1}^{N-1}
\frac{W_{m\iota}\,\Gamma_\text{abs}(\epsilon)}{\Gamma_\text{em}(\epsilon) + \Gamma_\text{esc}(\epsilon)[1-\cos(\pi{m}/N)]},\\
&W_{m\iota}\equiv\frac{1-(-1)^m}{N}\cot\frac{\pi{m}}{2N}\sin\frac{\pi{m}\,\iota}N.
\end{align}\end{subequations}
The largest weight is on the island $\iota=(N-1)/2$,:
\begin{equation}
W_{(N-1)/2}
=\frac2N\sin\frac{\pi{m}}{2}\cot\frac{\pi{m}}{2N}\cos\frac{\pi{m}}{2N},
\end{equation}
while $S_\text{qp}(-\omega_a)$ is determined by the spatial average
\begin{equation}
\sum_{\iota=1}^{N-1}\frac{W_{m\iota}}{N-1}
=\frac{1-(-1)^m}{N(N-1)}\,\cot^2\frac{\pi{m}}{2N}.
\end{equation}
Both these quantities decay with increasing $m$. They are equal to~1 for $N=3$, $m=1$ (the Kerr qubit of Ref.~\cite{Grimm2020}), and are close to~1 for $N=5$, $m=1$ (the dissipative qubit of Ref.~\cite{Lescanne2020}), while for $m=3$ they are much smaller.
Thus, we write
\begin{subequations}\label{eqs:Lambdas}
\begin{align}
&\tilde{f}(\epsilon\approx\omega_\text{d}) = \sqrt{\frac{\omega_\text{d}}{\epsilon-\omega_\text{d}}}\,
\frac{f(\epsilon-\omega_\text{d})}%
{\Lambda_\text{ph} + \Lambda_\text{esc}},\\
&\frac1{\Lambda_\text{ph}}\equiv\frac{315\,\zeta(5)}{8\pi}\,
\frac{E_J'|\varphi_{\text{d}}'|^2}{\Sigma{V}\omega_\text{d}'{}^4}\,\cos^2\frac{\varphi_{\text{bias}}'}{2},\\
&\frac1{\Lambda_\text{esc}}\equiv
\frac{1}{1-\cos(\pi/N)}\,
\frac{|\varphi_{\text{d}}'{}|^2/4}{\tan^2(\varphi_{\text{bias}}'/2) + 2\omega_\text{d}/\Delta},
\end{align}\end{subequations}
while for $\tilde{f}(\epsilon\approx\omega_a)$ we have a similar expression, differing by a substitution $\omega_\text{d}\to\omega_a$, $|\varphi_{\text{d}}'|^2\to|\varphi_a'|^2\langle\hat{a}^\dagger\hat{a}\rangle$. The coefficient $1/(\Lambda_\text{ph} + \Lambda_\text{esc})$ gives the fraction of the low-temperature quasiparticle population transferred to the higher energies around $\epsilon\approx\omega_\text{d}$ (recall that the number of quasiparticles involves the integration with the density of states, $\propto1/\sqrt\epsilon$). 
The dimensionless quantities $\Lambda_\text{ph}$ and $\Lambda_\text{esc}$, respectively, determine the relative efficiencies of the two cooling mechanisms, namely, the phonon emission and the quasiparticle escape to large islands.

Thus obtained correction to the distribution function contributes to the error rates via the quasiparticle structure factor. Namely, plugging $\tilde{f}(\epsilon\approx\omega_\text{d})$ into Eq.~(\ref{eq:Sqpgeneral}), we obtain $S_\text{qp}(-\omega_a) = S_\text{qp}(\omega_\text{d}-\omega_a)/(\Lambda_\text{ph} + \Lambda_\text{esc})$ for junctions connecting two small islands, and half of this value for junctions connecting a small and a large island (since hot quasiparticles are available only on the small island). We remind that the contribution of $\tilde{f}(\epsilon \approx \omega_a)$ cannot be included into the photon gain rate as discussed above (Sec. \ref{ssec:qualitativekinetics}).

The approximate values of the island volumes are $V\approx \SI{0.001}{\micro \cubic \meter}$
and $V\approx \SI{0.2}{\micro \cubic \meter}$
in Refs.~\cite{Grimm2020, Lescanne2020}, respectively~\cite{Grimm_Leghtas_private}.
Using $\Sigma = 0.3\times10^9\:\text{W}/(\text{m}^2\cdot\text{K}^5)$ for aluminum~\cite{Meschke2004}, and the same qubit parameters as in Secs.~\ref{sec:DerivationKerr},~\ref{sec:DerivationDissip}, we obtain the values of $\Lambda_\text{ph}$ and $\Lambda_\text{esc}$ listed in Table~\ref{tab:Lambda}. We see that for islands of such volume, the phonon emission does not prevent quasiparticle heating by photon absorption, and it is the exchange with large islands, rather than phonons, that maintains the thermal distribution. The found values of $\Lambda_\text{ph}$ and $\Lambda_\text{esc}$ for $\epsilon\approx\omega_\text{d}$ result in a contribution to $\kappa_+$ about $5-6$ times smaller than the one given in Eq.~(\ref{eq:kappa_plus}) and estimated in Secs.~\ref{sec:DerivationKerr},~\ref{sec:DerivationDissip}. Note, however, that the values of $\Lambda_{\rm ph},\Lambda_{\rm esc}$ are very much dependent on specifics of the structure, such that quasiparticle overheating cannot be {\it a priori} disregarded in any device.

\begin{table}
\begin{tabular}{|l|c|c|}
\hline & $\Lambda_\text{ph}$ & $\Lambda_\text{esc}$ \\
\hline Ref.~\cite{Grimm2020}, $\epsilon\approx\omega_\text{d}$ & 0.4 & $2\times10^3$\\
\hline Ref.~\cite{Grimm2020}, $\epsilon\approx\omega_a$ & $10^{-3}$ & 50\\
\hline Ref.~\cite{Lescanne2020}, $\epsilon\approx\omega_\text{d}$ & 600 & $10^4$\\
\hline Ref.~\cite{Lescanne2020}, $\epsilon\approx\omega_a$ & 20 & $10^3$\\
\hline
\end{tabular}
\caption{Values of the dimensionless parameters suppressing the population of excited quasiparticles, as defined in Eqs.~(\ref{eqs:Lambdas}).}
\label{tab:Lambda}
\end{table}

\section{Conclusions}
We have conducted a comprehensive analysis of the influence of Bogolyubov quasiparticles on Schr\"odinger cat qubits, whose operation intrinsically involves an external drive and, possibly, dissipation. Starting from the quasiparticle tunneling Hamiltonian, we derived microscopically the error dissipators appearing in the phenomenological master equation~(\ref{eq:masterequation}) and uncovered the limitations on its validity.
In particular, we found the single-photon loss rate to be similar to the relaxation rate of the excited state in undriven superconducting qubits. 
On the contrary, the single-photon gain rate in driven qubits is significantly enhanced with respect to the thermal rate in undriven qubits, since quasiparticles can absorb additional photons from the drive and transfer extra energy to the qubit. 
The pure dephasing rate, whose perturbative derivation results in a logarithmic divergence for conventional undriven qubits, behaves more regularly for cat qubits where most of the divergence is cured on the intrinsic time scales of the cat qubit. However, this regularization works only for transitions that start from the computational subspace of the cat qubit. In that case, the phenomenological master equation~(\ref{eq:masterequation}) can be written with logarithmic precision.

Like in conventional undriven superconducting qubits, quasiparticles thus constitute an intrinsic source of errors in both Kerr and dissipative cat qubits. Estimating the error rates for the existing cat qubit devices, we conclude that they are not yet at the stage where quasiparticle-induced errors represent their main limitation.

\acknowledgements

We thank A.~Bienfait, P.~Campagne-Ibarcq, M. Mirrahimi, and P.~Rouchon for helpful discussions.
We are also grateful to G.~Catelani for discussions and critical reading of the paper.

We also thank A.~Grimm and Z.~Leghtas for sharing details of their experiments.
We acknowledge funding from the Plan France 2030 through the project ANR-22-PETQ-0006.

\appendix

\section{Perturbations of the dissipative cat qubit}
\label{app:perturbations}


Here we analyze the slow dynamics of the dissipative cat qubit, perturbed by the error dissipators in Eq.~(\ref{eq:masterequation}). For this, it is convenient to introduce the eigenstates $|\psi_{l\sigma}\rangle$ and the eigenvalues $\mu_{l\sigma}$ of the Kerr Hamiltonian, such that  $(\hat{a}^\dagger{}^2-\alpha^2)(\hat{a}^2-\alpha^2)|\psi_{l\sigma}\rangle=\mu_{l\sigma}|\psi_{l\sigma}\rangle$, classified by the parity $\sigma=(-1)^{\hat{a}^\dagger\hat{a}}$ and an integer $l\geq0$. The two eigenvectors, corresponding to $\mu_{0\pm}=0$, are the cat states $|\psi_{0\pm}\rangle\equiv|\mathcal{C}_\alpha^\pm\rangle$. Then, besides the obvious four right eigenvectors $\hat\varrho_0^{\sigma\sigma'}=|\psi_{0\sigma}\rangle\langle\psi_{0\sigma'}|$ of $\mathcal{L}_0=\kappa_2\mathcal{D}[\hat{a}^2-\alpha^2]$ with the eigenvalue $\lambda=0$, we can construct additional right eigenvectors (still not forming a complete set), noting that
\begin{subequations}\label{eqs:Lindbladian_eigenvectorsl00l}
\begin{align}
&\mathcal{L}_0|\psi_{l\sigma}\rangle\langle\psi_{0\sigma'}| =
-\frac{\kappa_2\mu_{l\sigma}}2\,|\psi_{l\sigma}\rangle\langle\psi_{0\sigma'}|,\\
&\mathcal{L}_0|\psi_{0\sigma}\rangle\langle\psi_{l'\sigma'}| =
-\frac{\kappa_2\mu_{l'\sigma'}}2\,|\psi_{0\sigma}\rangle\langle\psi_{l'\sigma'}|.
\end{align}\end{subequations}
The four left eigenvectors $\hat\varsigma_0^{\sigma\sigma'}$ of $\mathcal{L}_0$, corresponding to the zero eigenvalues, called invariant operators since $(\partial/\partial{t})\Tr\{\hat\varsigma_0^{\sigma\sigma'}\hat\rho\}=0$, are also known~\cite{Guillaud2023} and given by (up to a normalization):
\begin{subequations}\begin{align}
&\hat\varsigma_0^{\sigma\sigma}=\sum_{l=0}^\infty|\psi_{l\sigma}\rangle\langle\psi_{l\sigma}|,\\
&\hat\varsigma_0^{+-}=
\sum_{n,m=0}^\infty\frac{(-1)^{n-m}I_{n-m}(\alpha^2)}{2n+1-2m}\,
\frac{\hat{a}^\dagger{}^{2n+1}|0\rangle\langle0|\hat{a}^{2m}}{(2n)!!\,(2m)!!}\nonumber\\
&\qquad{}=(\hat\varsigma_0^{-+})^\dagger,
\end{align}\end{subequations}
as can be found by explicitly acting from the left, $\hat{a}^\dagger{}^{2n+1}|0\rangle\langle0|\hat{a}^{2m}\mathcal{L}_0$, and using the recursive relation for the modified Bessel function $I_n(z)$: 
\begin{subequations}\begin{align}
&\frac{2n}z\,I_n(z)=I_{n-1}(z)-I_{n+1}(z),\\
&I_n(z)=\int_{-\pi}^\pi\frac{d\phi}{2\pi}\,e^{z\cos\phi+in\phi}.
\end{align}\end{subequations}
Using $1/(2n-2m+1)=(i/2)\int_0^\pi{e}^{-(2n-2m+1)i\theta}\,d\theta$ and $\int_0^{\pi/2}I_0(z\cos\theta)\,z\cos\theta\,d\theta=\sinh{z}$, we calculate the matrix elements between the coherent states $|{\pm\alpha}\rangle$:
\begin{align}
&\langle\sigma\alpha|\hat\varsigma_0^{+-}|\sigma'\alpha\rangle=\nonumber\\
{}&{}=\sigma{e}^{-\alpha^2}
\sum_{n,m=0}^\infty\frac{(-1)^{n-m}\alpha^{2n+2m+1}I_{n-m}(\alpha^2)}{(2n-2m+1)\,(2n)!!\,(2m)!!}
{}\nonumber\\
{}&{}=\frac{i\sigma}2\,\alpha e^{-\alpha^2}
\int_0^\pi{d}\theta\int_{-\pi}^\pi\frac{d\phi}{2\pi}\,
e^{-i\theta-2\alpha^2\sin\phi\sin\theta}
\nonumber\\
{}&{}=
\sigma{e}^{-\alpha^2}\,\frac{\sinh2\alpha^2}{2\alpha}
\end{align}
(independent of $\sigma'$), which gives
\begin{equation}\label{eq:normalizationfactor}
\langle\mathcal{C}_\alpha^-|\hat\varsigma_0^{+-}|\mathcal{C}_\alpha^+\rangle
= \sqrt{\frac{\sinh2\alpha^2}{2\alpha^2}} =
\langle\mathcal{C}_\alpha^+|\hat\varsigma_0^{-+}|\mathcal{C}_\alpha^-\rangle.
\end{equation}
This factor has to be included when projecting on the zero subspace, since the left eigenvectors (\ref{eqs:Lindbladian_eigenvectorsl00l}) are not normalized.


Similarly to the previous calculation, using $\int_0^{\pi/2}I_1(2z\cos\theta)\,d\theta=(\sinh^2z)/z$, we calculate
\begin{align}
&\langle\sigma\alpha|\hat{a}\hat\varsigma_0^{+-}\hat{a}^\dagger|\sigma'\alpha\rangle={}\nonumber\\
&{}=\frac{\sigma'{e}^{-\alpha^2}}{\alpha}
\sum_{n,m=0}^\infty\frac{(-1)^{n-m}\alpha^{2n+2m}I_{n-m}(\alpha^2)}{(2n-2m+1)\,(2n)!!\,(2m)!!}\,(2n+1)2m
{}\nonumber\\ {}&{}=\frac{i\sigma'}2\,\alpha e^{-\alpha^2}
\int\limits_0^\pi{d}\theta\int\limits_{-\pi}^\pi\frac{d\phi}{2\pi}\,
e^{-2\alpha^2\sin\phi\sin\theta}
\left(\alpha^2e^{-i\theta}-e^{-i\phi}\right)
{}\nonumber\\ {}&{}=\sigma'\alpha e^{-\alpha^2}\int\limits_0^{\pi/2}
\left[I_0(2\alpha^2\cos\theta)\,\alpha^2\cos\theta+I_1(2\alpha^2\cos\theta)\right]d\theta,
\end{align}
which gives
\begin{align}
\langle\mathcal{C}^+_\alpha|\hat{a}\hat\varsigma_0^{+-}\hat{a}^\dagger|\mathcal{C}^-_\alpha\rangle=
{}&{}
\sqrt{\frac{\sinh2\alpha^2}{2\alpha^2}}\left(\alpha^2+\tanh\alpha^2\right).
\end{align}
These results enable us to express the projections of various Lindbladian perturbations $\mathcal{L}_1$ on the zero subspace of $\mathcal{L}_0$ according to
\begin{equation}
\mathcal{P}_\|\mathcal{L}_1|\mathcal{C}_\alpha^\sigma\rangle\langle\mathcal{C}_\alpha^{\sigma'}|
=\sum_{\sigma_1,\sigma_1'}
\frac{\Tr\{\varsigma_0^{\sigma_1\sigma_1'}\mathcal{L}_1|\mathcal{C}_\alpha^\sigma\rangle\langle\mathcal{C}_\alpha^{\sigma'}|\}}{\langle\mathcal{C}_\alpha^{\sigma_1'}|\varsigma_0^{\sigma_1\sigma_1'}|\mathcal{C}_\alpha^{\sigma_1}\rangle}
|\mathcal{C}_\alpha^{\sigma_1}\rangle\langle\mathcal{C}_\alpha^{\sigma_1'}|.
\end{equation}
Noting that
\begin{equation}\label{eq:hpm}
\hat{a}|\mathcal{C}_\alpha^\sigma\rangle=\alpha\sqrt{h_\sigma}|\mathcal{C}_\alpha^{-\sigma}\rangle,
\quad
h_+\equiv\tanh\alpha^2,\quad h_-\equiv\coth\alpha^2,
\end{equation}
we find
\begin{subequations}\begin{align}
\mathcal{P}_\|\mathcal{D}[\hat{a}]
|\mathcal{C}_\alpha^\sigma\rangle\langle\mathcal{C}_\alpha^{\sigma'}|= {}&{}
\alpha^2\sqrt{h_\sigma{h}_{\sigma'}}
|\mathcal{C}_\alpha^{-\sigma}\rangle\langle\mathcal{C}_\alpha^{-\sigma'}|\nonumber\\
{}&{}-\alpha^2\,\frac{h_\sigma+{h}_{\sigma'}}2\,
|\mathcal{C}_\alpha^\sigma\rangle\langle\mathcal{C}_\alpha^{\sigma'}|,\\
\mathcal{P}_\|\mathcal{D}[\hat{a}^\dagger]
|\mathcal{C}_\alpha^\sigma\rangle\langle\mathcal{C}_\alpha^{\sigma'}|={}&{}
\delta_{\sigma\sigma'}(1+\alpha^2h_\sigma)
|\mathcal{C}_\alpha^{-\sigma}\rangle\langle\mathcal{C}_\alpha^{-\sigma'}|\nonumber\\
{}&{}+\delta_{\sigma,-\sigma'}(\alpha^2+h_+)
|\mathcal{C}_\alpha^{-\sigma}\rangle\langle\mathcal{C}_\alpha^{-\sigma'}|\nonumber\\
{}&{}
-\left(1+\alpha^2\,\frac{h_\sigma+{h}_{\sigma'}}2\right)
|\mathcal{C}_\alpha^\sigma\rangle\langle\mathcal{C}_\alpha^{\sigma'}|,\\
\mathcal{P}_\|\mathcal{D}[\hat{a}^\dagger\hat{a}]
|\mathcal{C}_\alpha^\sigma\rangle\langle\mathcal{C}_\alpha^{\sigma'}|={}&{}
-\frac{\delta_{\sigma,-\sigma'}\alpha^2}{\sinh2\alpha^2}
|\mathcal{C}_\alpha^\sigma\rangle\langle\mathcal{C}_\alpha^{\sigma'}|.
\end{align}\end{subequations}
These $4\times4$ matrices have a $2\times2$ block structure, since all perturbations conserve the product of the left and right parities; their eigenvalues are found straightforwardly, leading to Eqs.~(\ref{eqs:lambdaDissip}) in the limit $\alpha^2\gg1$.
Note that since the perturbation $\mathcal{D}[\hat{a}^\dagger\hat{a}]$ conserves the left and the right parity separately, it produces a zero eigenvalue for both $|\mathcal{C}_\alpha^+\rangle\langle\mathcal{C}_\alpha^+|$ and $|\mathcal{C}_\alpha^-\rangle\langle\mathcal{C}_\alpha^-|$.

\begin{figure}
\includegraphics[width=0.3\textwidth]{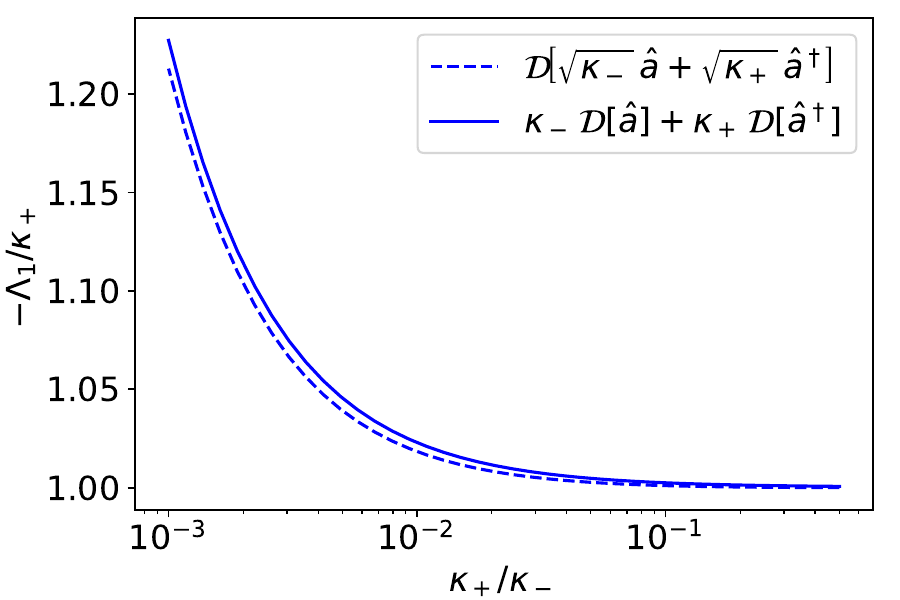}
\includegraphics[width=0.3\textwidth]{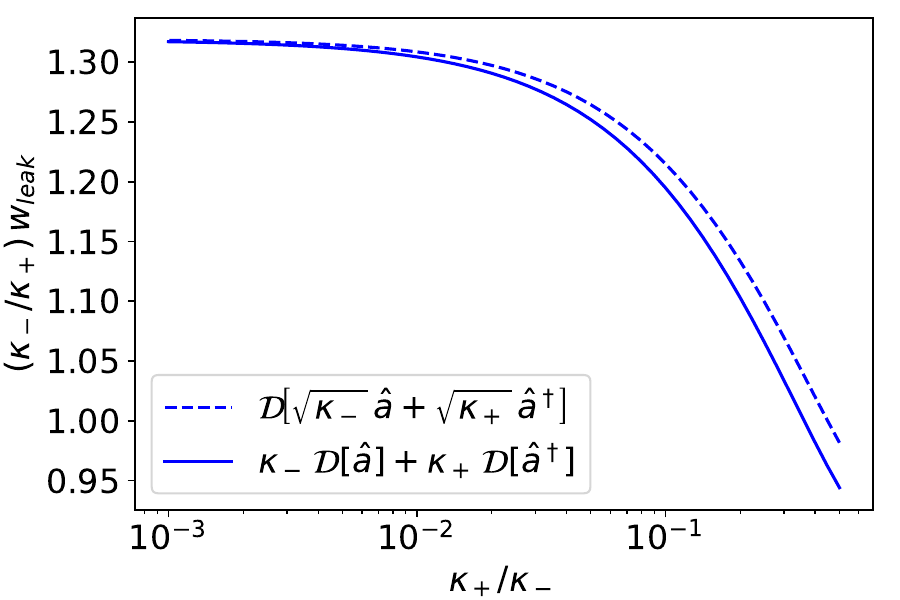}
\caption{Comparison of the results obtained from the phenomenological master equation~(\ref{eq:masterequation}) with $\alpha^2=2.5$, $\kappa_-/K=10^{-3}$, $\kappa_\phi=0$ (solid lines) and from the master equation where the sum of the dissipators $\kappa_-\mathcal{D}[\hat{a}]+\kappa_+\mathcal{D}[\hat{a}^\dagger]$ is replaced by a single coherent dissipator $\mathcal{D}\left[\sqrt{\kappa_-}\,\hat{a}+\sqrt{\kappa_+}\,\hat{a}^\dagger\right]$ (dashed lines). In the upper panel, we plot the first non-zero eigenvalue $\lambda_1$, and on the lower panel the leakage probability $w_\text{leak}$ (the probability to be outside the computational subspace) in the stationary state, defined in Eq.~(\ref{eq:werr}).}
\label{fig:plot_lambda1_vs_kp}
\end{figure}

\section{Numerical results for the phenomenological master equation}
\label{app:numerics}

Here we address the issue of frequency matching raised in Secs.~\ref{ssec:quasiparticle-induced} and~\ref{sec:DerivationKerr}.  We numerically solve Eq.~(\ref{eq:masterequation}), as well as its counterpart with the sum of the photon loss (gain) dissipators  $\kappa_-\mathcal{D}[\hat{a}]+\kappa_+\mathcal{D}[\hat{a}^\dagger]$ replaced by a single coherent dissipator $\mathcal{D}\left[\sqrt{\kappa_-}\,\hat{a}+\sqrt{\kappa_+}\,\hat{a}^\dagger\right]$.
The interesting quantities to compare are (i)~the first nonzero eigenvalue $\lambda_1$, and (ii)~the probability $w_\text{leak}$ to be outside the computational space, as defined by Eq.~(\ref{eq:werr}), evaluated in the stationary state of the full Lindbladian.

For Eq.~(\ref{eq:masterequation}), the first nonzero eigenvalue as a function of~$\alpha^2$ exhibits a plateau which begins approximately when the leakage terms balance the photon loss in Eq.~\eqref{eq:lambda_xKerrProjected}, $2\kappa_-\alpha^2e^{-4\alpha^2}\sim\kappa_++\kappa_\phi\alpha^2$~\cite{Gautier2022, Frattini2022}.
From a practical point of view, it is convenient to work with $\alpha^2$ in the beginning of the plateau, since for larger $\alpha^2$ other error rates increase. Thus, we choose $\alpha^2=2.5$ used in the experiment~\cite{Grimm2020}.

For such moderate values of $\alpha^2$, it is possible to use brute force, representing the Lindbladian as a matrix in the basis $|n\rangle\langle{n}'|$, built from the Fock states $|n\rangle$ of the harmonic oscillator, $\hat{a}^\dagger\hat{a}|n\rangle=n|n\rangle$. For $\alpha^2=2.5$, truncation at $n=30$ is sufficient to obtain a precision exceeding the thickness of the curves in the figures.
 
Using the brute force diagonalization of the full Lindbladian, we calculate the first nonzero eigenvalue $\lambda_1$, as well as the probability $w_\text{leak}$ to be outside the computational space in the stationary state, Eq.~(\ref{eq:werr}).
For both quantities, the two solutions are very close,  as shown in Fig.~\ref{fig:plot_lambda1_vs_kp}, so we conclude that the frequency-matching issue can be ignored for practical purposes.

\bibliography{references}

\end{document}